\documentclass{emulateapj}

\usepackage{newtxtext,newtxmath}
\usepackage[T1]{fontenc}
\usepackage{ae,aecompl}

\usepackage{xspace}
\usepackage{graphics,graphicx}
\usepackage{natbib}
\usepackage{multirow}
\usepackage{color}

\newcommand{\nH}{$N_{\rm H}$\xspace}

\newcommand{\Msun}{$M_{\odot}$\xspace}

\newcommand{\chisq}{$\chi^2$\xspace}

\newcommand{\Chandra}{{\it Chandra}\xspace}

\newcommand{\Swift}{{\it Swift}\xspace}
\newcommand{\XMM}{{\it XMM-Newton}\xspace}
\newcommand{\nustar}{{\it NuSTAR}\xspace}
\newcommand{\lum}{erg s$^{-1}$\xspace}
\newcommand{\flux}{erg s$^{-1}$ cm$^{-2}$\xspace}

\shorttitle{Geometric Beaming in NGC 300 ULX-1}
\shortauthors{Binder et al.}

\begin{document}

\title{No Strong Geometric Beaming in the Ultraluminous Neutron Star Binary NGC 300 ULX-1 (SN 2010da) from Swift and Gemini}

\author{B. Binder\altaffilmark{1}, E. M. Levesque\altaffilmark{2}, T. Dorn-Wallenstein\altaffilmark{2}
}

\altaffiltext{1}{Department of Physics \& Astronomy, California State Polytechnic University, 3801 W. Temple Avenue, Pomona, CA 91768 }
\altaffiltext{2}{University of Washington, Department of Astronomy, Box 351580, Seattle, WA 98195}

\begin{abstract}
We have obtained near-simultaneous \Swift/XRT imaging and Gemini GMOS spectroscopy for the ultraluminous X-ray source (ULX) NGC~300 ULX-1 (formerly designated SN~2010da). The observed X-ray emission is consistent with an inhomogeneous wind that partially obscures a central, bright inner accretion disk. We simultaneously fit eleven 0.3-10 keV spectra obtained over a $\sim$1 year time period (2016 April to 2017 July) using the same partial covering model, and find that although the covering fraction varies significantly (from 78\% to consistent with 0\%), the unabsorbed luminosity remains essentially constant across all observations ($2-6\times10^{39}$ \lum). A relatively high 0.3-10 keV fractional variability amplitude ($F_{\rm var}$) of $\sim$30\% is observed in all eleven observations. Optical spectra from Gemini exhibit numerous emission lines (e.g., H$\alpha$, H$\beta$, He~II $\lambda$4686) which suggest that the neutron star primary is photoionizing material in the immediate vicinity of the binary. We compare the He~II $\lambda$4686 line luminosity ($\sim7-9\times10^{35}$ \lum) to the contemporaneous soft X-ray emission and find the X-ray emission is broadly consistent with the observed He~II line luminosity. The combination of our X-ray observations and optical spectroscopy suggest that geometric beaming effects in the ULX-1 system are minimal, making ULX-1 one of only a few bona fide ULXs to be powered by accretion onto a neutron star. 
\end{abstract}


\section{Introduction}
It is now well-established that the optical transient SN~2010da \citep{Monard10,Immler+10} in NGC~300 was a supernova ``impostor'' \citep{Chornock+10,EliasRosa+10a,EliasRosa+10b}, and likely a young \citep[$<$5 Myr;][]{Binder+11,Binder+16} high-mass X-ray binary (HMXB). Several multiwavelength studies (from X-ray to infrared) have been conducted to examine the nature of the binary components \citep{Binder+16,Villar+16,Lau+16}, and recently deep \XMM+\nustar observations have confirmed the neutron star origin of the X-ray emission \citep{Carpano+18,Walton+18,Kosec+18}. The most likely scenario is that SN~2010da hosts a neutron star primary with a spin period $P_s\sim31.6$ s and $\dot{P}=-5.4\times10^{-7}$ s$^{-1}$ \citep{Carpano+18} and a supergiant companion \citep{Lau+16}, possibly a yellow- or red-supergiant (RSG) entering a blue-loop phase of its evolution \citep{Villar+16}. Optical spectroscopy \citep{Chornock+10,EliasRosa+10b,Chornock+11,Villar+16} has revealed complex environment. Much of the dust that enshrouded the progenitor system \citep{Khan+10} was destroyed in the initial optical/X-ray outburst \citep{Brown10,Prieto+10}, and recent observations suggest that new dust is actively forming from the supergiant donor star \citep{Lau+16}. The high X-ray luminosities ($\sim10^{39}$ \lum) recently observed in the XMM and \nustar have earned SN~2010da a new designation: NGC~300 ULX-1.

Ultraluminous X-ray sources (ULXs) are off-nuclear X-ray sources with luminosities exceeding the Eddington limit of a 10 \Msun black hole in the 0.3-10 keV energy band \citep[for a review, see][]{Feng+Soria11}. The extreme X-ray luminosities are expected to arise from `supercritical' accretion \citep{King09,Bachetti+14}, where large scale-height, optically thick winds are radiatively driven to large radii, revealing a hot accretion disk \citep{King04,Poutanen+07}. Inclination angle, therefore, plays a major role in the appearance of ULXs \citep{Middleton+11a,Middleton+14a,Sutton+13}, and beaming effects may amplify the X-ray emission \citep[particularly in the hard band][]{King09,Poutanen+07}. Clumps and inhomogeneities in the winds can imprint variability in the X-ray emission, which is likewise dependent on inclination angle \citep{Takeuchi+13,Takeuchi+14,Middleton+11a,Heil+09,Sutton+13}.

We present the results of recent monitoring campaign of NGC~300 ULX-1/SN~2010da (hereafter, ULX-1) by the Neil Gehrels \Swift Observatory (hereafter, \Swift) combined with ground-based optical spectroscopy with the Gemini Observatory. Throughout this work, we assume a distance to NGC~300 of 2.0 Mpc \citep[][ corresponding to a redshift $z=0.00047$]{Dalcanton+09}, and a column of neutral absorption fixed at the Galactic column \citep[\nH = 4.09$\times10^{20}$ cm$^{-2}$;][]{Kalberla+05}.

\section{Observations}
	\subsection{\Swift/XRT}
\Swift/XRT \citep{Burrows+04} observations were obtained from 2016 April 14 through 2017 April 22 through the \Swift archive \citep{Evans+09}, and our team acquired three new target-of-opportunity observations on 2017 July 2, 6, and 12 to coincide with our Gemini observations. All observations were performed in Photon Counting mode \citep{Hill+04}. The data were produced using standard procedures (\texttt{XRTPIPELINE}) using the \texttt{FTOOLS} in the \texttt{HEASOFT} package (v 6.17).

\begin{figure}
\centering
\includegraphics[width=0.85\linewidth,clip=true,trim=0.8cm 0.8cm 0cm 0cm]{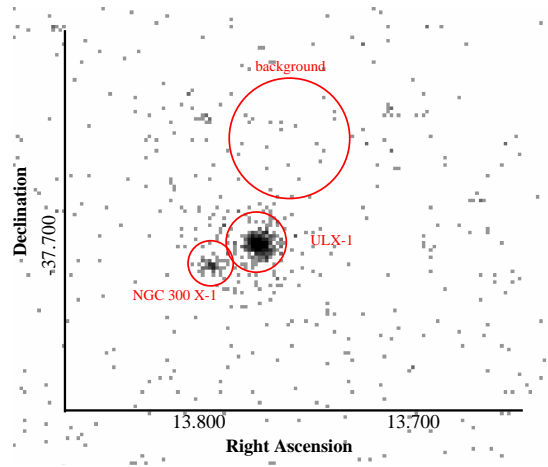}	\\
\caption{The 0.3-10 keV \Swift/XRT image of ULX-1 (SN~2010da) from ObsID 00049834010. The source and background extraction regions are shown, as well as the location of the nearby X-ray source NGC~300 X-1.}
\label{figure:Xray_image}
\end{figure}

The location of the X-ray source corresponding to ULX-1 has been measured with high precision by \Chandra, with a right ascension of 00:55:04.9 and a declination of -37:41:44.0. Spectra were extracted in each observations where an X-ray source was detected near these coordinates with a signal-to-noise $\geq$3; we used a circular aperture with a 40\arcsec\xspace radius centered on the \Chandra position of ULX-1. The background was extracted from a circular aperture (radius of 80\arcsec) offset from the ULX-1 position in a region with no obvious contaminating X-ray sources. Figure~\ref{figure:Xray_image} shows an example \Swift/XRT image, with our extraction regions superimposed.  Ancillary response files were generated for each spectrum with \texttt{XRTMKARF} using the spectral redistribution matrix v015\footnote{See \url{http://heasarc.gsfc.nasa.gov/docs/heasarc/caldb/swift}}.

	\subsection{Gemini Spectroscopy}
We obtained time-resolved optical spectra of ULX-1 using the Gemini Multi-Object Spectrograph on Gemini South \citep{Hook+04,Gimeno+16}. The observations were carried out in queue mode during June and July of 2017 and were originally designed to be executed in three separate visits, each with complete wavelength coverage, spaced 5 and 10 days apart respectively. Unfortunately, poor weather at Gemini prevented this execution, and we instead obtained four different spectra with full or partial wavelength coverage spanning a 42-day period. Our final set of observations is summarized in Table~\ref{table:Obs}.

\begin{table}
\centering
\caption{Summary of Gemini Observations}
\begin{tabular}{ccccc}
\hline \hline
Date 	& \multirow{2}{*}{Grating}	& $\lambda_C$	& Exp. Time &Resolution \\
(UT)		&		& (\AA)		& (s)		& (\AA)	\\
(1)			& (2)		& (3)					& (4)	& (5)		\\
\hline
2017 Jun 19 & B600 & 5000 & 2 $\times$ 600 &3\\
 & B600 & 5050 & 2 $\times$ 600 &3 \\
 & R400 & 7000 & 2 $\times$ 600 &4 \\ 
2017 Jul 02 & B600 & 5000 & 2 $\times$ 600 &3 \\
 & B600 & 5050 & 2 $\times$ 600 &3 \\
 & R400 & 7000 & 2 $\times$ 600 &4 \\
 & R400 & 7050 & 2 $\times$ 600 &4 \\ 
2017 Jul 18 & R400 & 7000 & 2 $\times$ 600 &4 \\
 & R400 & 7050 & 2 $\times$ 600 &4 \\ 
2017 Jul 30 & B600 & 5000 & 2 $\times$ 600 &3 \\
 & B600 & 5050 & 2 $\times$ 600 &3 \\
\hline \hline
\end{tabular}
\label{table:Obs}
\end{table}

\begin{figure*}
\centering
\includegraphics[width=0.75\linewidth,clip=true,trim=0cm 2.5cm 0cm 3cm]{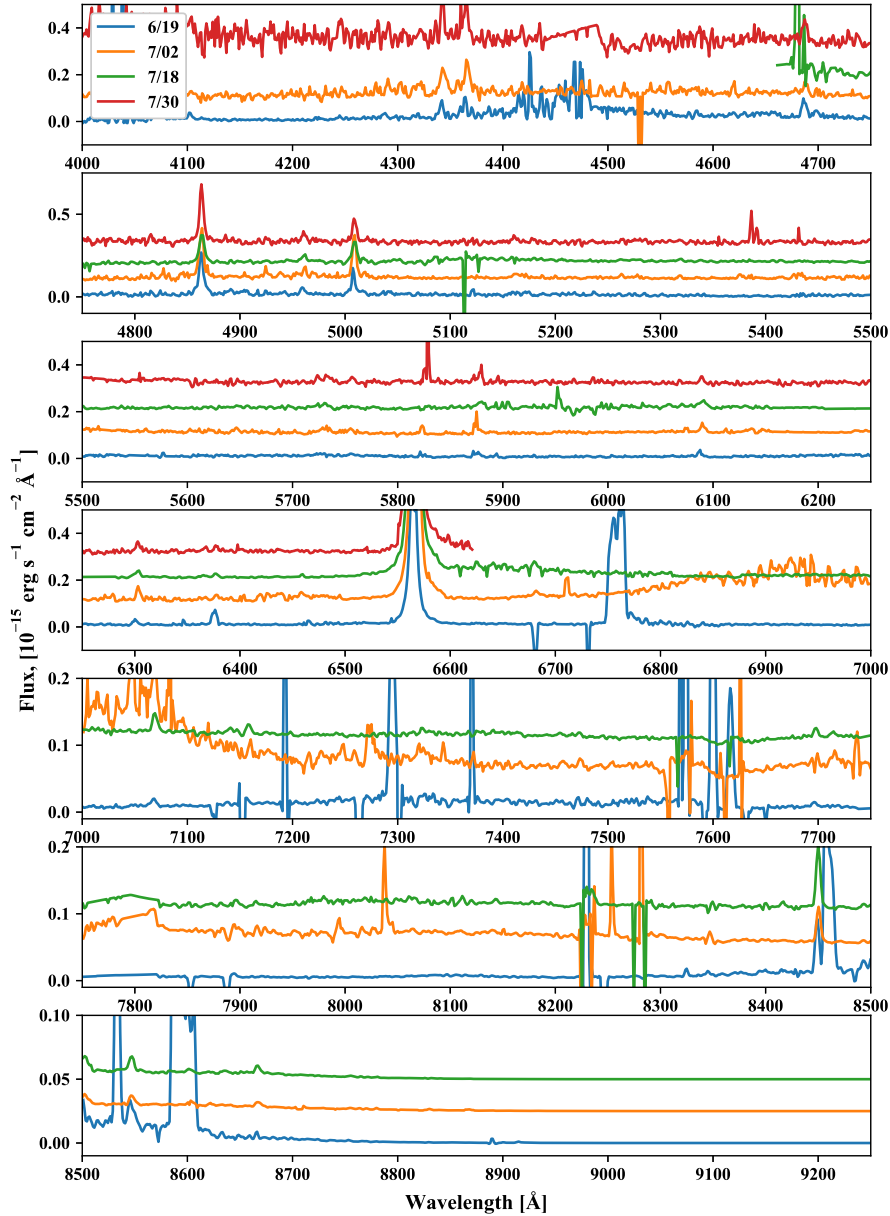}	\\
\caption{The four Gemini GMOS spectra of ULX-1 (SN~2010da). Flux levels have an arbitrary offset for clarity.}
\label{figure:Gemini_spectra}
\end{figure*}

Our spectra were taken with the 0.5$''\times$330$''$ slit, using the B600 grating centered at 5000\AA\ and the R400 grating (with the GG455 blocking filter) centered at 7000\AA\ for a total wavelength coverage of $\sim$3600\AA\ to 9000\AA; the grating centers shifted by +50\AA\ for half of our observations to avoid the detector gaps. Image quality during these observations was $\ge$85$\%$ \footnote[2]{Our image quality criteria correspond to a point source FWHM of $\le1.05''$ in the $r$-band; see \url{http://www.gemini.edu/sciops/telescopes-and-sites/observing-condition-constraints\#ImageQuality} for a complete discussion of Gemini image quality criteria}, and the data were taken at a mean airmass of 1.1. Standard quartz lamp and CuAr arc lamp observations were taken for flat-field and wavelength calibrations.

The data were reduced using the {\tt gemini} package in {\tt IRAF} \citep{Gemini}. Data read out through the 12 amplifiers on the three GMOS-S chips were mosaiced together with the {\tt gprepare} task. A bias image and flat field images for each grating were created from the {\tt gbias} and {\tt gsflat} tasks, and the data were reduced with {\tt gsreduce}. Two-dimensional wavelength solutions in each grating were created with {\tt gswavelength} and applied with {\tt gstransform}. We used {\tt gsskysub} on the two dimensional images to account for telluric features in the data. One-dimensional spectra were traced, extracted, and background-subtracted with {\tt gsextract}. We also observed the spectrophotometric standard star Hiltner 600 \citep[aka H600/HD 289002,][]{Baldwin+Stone84} with the same instrument setup on 2017 March 3rd -- unfortunately we were unable to obtain observations of H600 on the same night as our observations of ULX-1 due to poor weather. We derive the sensitivity function for each grating using {\tt gsstandard} and applied it to the data with {\tt gscalibrate}. Finally, the data from individual nights were combined using {\tt gemscombine}. 

The resulting flux-calibrated GMOS spectra are shown in Figure~\ref{figure:Gemini_spectra}, and line identifications and fluxes \citep[corrected for a foreground reddening of E(B-V)=0.011;][]{Schlafly+11} are listed in Table~\ref{table:line_IDs}. The wavelengths in both Figure~\ref{figure:Gemini_spectra} and Table~\ref{table:line_IDs} are in the observer frame \citep[as was done in][]{Villar+16}. The redshift to NGC~300 is very small ($\sim$141 km s$^{-1}$), and it is thus impossible to clarify whether the observed shifts in the lines relative to the rest wavelength can be attributed to the velocity of NGC~300 itself, the motion of ULX-1 within NGC~300, or the velocity of the emission source within the system.

\begin{table*}\setlength{\tabcolsep}{2.2pt}
\centering
\scriptsize
\caption{Optical Line Identifications}
\begin{tabular}{cccccccccccccc}
\hline \hline
Date		& Line ID	& Shape$^a$	&$\lambda_{\rm c}^b$	& Flux (10$^{-15}$ erg			& FWHM 			& Eq. Width	&& Line ID	& Shape$^a$	&$\lambda_{\rm c}^b$	& Flux (10$^{-15}$ erg 			& FWHM 			& Eq. Width	\\
		&		&			& (\AA)			& s$^{-1}$ cm$^{-2}$ \AA$^{-1}$)	& (km s$^{-1}$)		& (\AA)		&&			&			& (\AA)			& s$^{-1}$ cm$^{-2}$ \AA$^{-1}$)	& (km s$^{-1}$)		& (\AA)		\\
(1)		& (2)		& (3)			& (4)				& (5)							& (6)				& (7)			&& (8)		& (9)			& (10)			& (11)						& (12)			& (13)		\\
\hline
June 19	& \multirow{3}{*}{H$\delta$}	& G		& 4102.44		& 0.099	& 231	& -7.346	&& \multirow{3}{*}{H$\gamma$}		& G		& 4341.65		& 0.338	& 305	& -25.79	\\
July 2	&						& G		& 4103.80		& 0.134	& 323	& -8.034	&&							& G		& 4343.46		& 0.863	& 515	& -44.54	\\
July 30	&						&		& ...			& ...		& ...		& ...		&&							& G		& 4342.52		& 0.756	& 211	& -13.22	\\
\hline
June 19	& [O~III]+					& G		& 4364.25		& 0.243	& 185	& -8.522	&& \multirow{3}{*}{He~II}			& G		& 4686.71		& 0.430	& 338	& -21.2	\\
July 2	& [Fe~IX]+				& G		& 4366.16		& 0.688	& 354	& -21.89	&&							& G		& 4688.15		& 0.318	& 342	& -24.3	\\
July 30	& Fe~II					& G 		& 4364.34		& 0.640	& 149 	& -6.623	&&							& G		& 4687.07		& 0.587	& 355	& -15.4	\\
\hline
June 19	& \multirow{4}{*}{H$\beta$}	&  L		& 4863.16		& 1.967	& 280	& -205.2	&& \multirow{4}{*}{[O~III]}			& G		& 4959.57		& 0.238	& 296	& -16.86	\\
July 2	&						&  L		& 4863.45		& 2.103	& 253	& -109.5	&&							& G		& 4961.73		& 0.207	& 185	& -9.865	\\
July 18	&						&  L		& 4863.84		& 1.479	& 293	& -306.5	&&							& G		& 4960.89		& 0.207	& 324	& -16.14	\\
July 30	& 						&  L		& 4863.38		& 2.333	& 240	& -76.2	&&							& G		& 4960.88		& 0.334	& 268	& -6.50	\\
\hline
June 19	& \multirow{4}{*}{[O~III]}		& L		& 5007.77		& 0.875	& 186 	& -73.3	&& 							& G		& 5755.06		& 0.038 	& 251	& -3.583	\\
July 2	&						& G		& 5008.8		& 1.196	& 253	& -54.24	&& [N~II]+						& G 		& 5755.36		& 0.104 	& 243	& -11.98	\\
July 18	&						& G		& 5008.86		& 0.837	& 350	& -461.8	&& [Fe~II]						&		& ...			& ...		& ...		& ...	\\
July 30	& 						& G 		& 5008.93		& 0.682 	& 249	& -12.94	&&							& G		& 5756.86		& 0.018 	& 305	& -8.587	\\
\hline
June 19	& \multirow{4}{*}{He~I}		& c		& ...			& ...		& ...		& ...		&& \multirow{4}{*}{[Fe~VIII]}		& G		& 6087.1		& 0.094 	& 171	& -11.18	\\
July 2	& 						& c		& ...			& ...		& ...		& ...		&& 							& G		& 6090.11		& 0.174 	& 245 	& -13.54	\\
July 18	& 						& G		& 5879.01		& 0.293 	& 483	& -30.22	&& 							& G		& 6090.45		& 0.301 	& 423	& -14.00	\\
July 30	& 						& c		& ...			& ...		& ...		& ...		&& 							& G		& 6089.17		& 0.150  	& 225	& -8.665	\\
\hline
June 19	& \multirow{4}{*}{[O~I]}		& G		& 6300.33		& 0.147 	& 285	& -26.58	&& \multirow{4}{*}{[Fe~X]}			& G		& 6375.71		& 0.360 	& 246	& -34.98	\\
July 2	& 						& G		& 6303.35		& 0.271 	& 202	& -19.19	&& 							& G		& 6377.66		& 0.115 	& 255	& -8.165	\\
July 18	& 						& G		& 6302.66		& 0.151 	& 278	& -10.6	&& 							& G		& 6377.09		& 0.085 	& 211	& -6.445 \\
July 30	& 						& G 		& 6302.91		& 0.181 	& 194	& -6.267	&& 							& G		& 6377.24		& 0.114 	& 179	& -4.157	\\
\hline
June 19	& \multirow{4}{*}{H$\alpha$}	& L		& 6564.57		& 7.846 	& 343	& -453.7	&& \multirow{4}{*}{He~I}			& G		& 7067.89		& 0.080 	& 304	& -11.14	\\
July 2	& 						& L		& 6566.56		& 19.52 	& 299	& -961.7	&& 							& G		& 7068.98		& 0.377 	& 219	& -4.52	\\
July 18	& 						& L		& 6566.17		& 26.43 	& 298	& -696.2	&& 							& G		& 7069.2		& 0.167 	& 235	& -8.247	\\
July 30	& 						& L		& 6565.8		& 31.38 	& 257 	& -905.7	&& 							& 		& ...			& ...		& ...		& ...	\\
\hline
June 19	& \multirow{2}{*}{[Ca~II]}		& 		& ...			& ...		& ...		& ...		&& \multirow{2}{*}{[Fe~XI]}		& G		& 7894.32		& 0.033 	& 180	& -5.371	\\
July 18	& 						& G		& 7323.15		& 0.033 	& 140	& -1.961	&& 							& G		& 7896.61		& 0.082 	& 172	& -4.518	\\
\hline
June 19	& \multirow{3}{*}{O~I}		& G		& 8448.95		& 0.341 	& 151	& -22.66	&& Ca~II+						& G		& 8500.65		& 0.051 	& 114	& -3.931	\\
July 2	& 						& G		& 8450.46		& 0.247 	& 157	& -34.12	&& Pa~8500					& G		& 8502.96		& 0.045 	& 242	& -9.288	\\
July 18	& 						& G		& 8450.11		& 0.452 	& 157	& -39.83	&& 							& G		& 8503.0		& 0.067 	& 252	& -8.854	\\
\hline
June 19	& Ca~II+					& G		& 8546.89		& 0.154 	& 241	& -8.499	&& Ca~II+						& G		& 8665.86		& 0.018 	&100		& -5.757	\\
July 2	& Pa~8544				& G 		& 8547.34		& 0.051 	& 218	& -11.27	&& Pa~8664					& G		& 8667.63		& 0.025 	& 235	& -8.323	\\
July 18	& 						& G		& 8546.9		& 0.069 	& 199	& -12.06	&& 							& G		& 8666.86		& 0.033 	& 178	& -7.597	\\
\hline
June 19	& \multirow{3}{*}{Pa~8753}	& G		& 8753.51		& 0.004 	& 167	& -5.189	&& \multirow{3}{*}{Pa~8753}		& 		& ...			& ...		& ...		& ...	\\
July 2	& 						& G		& 8754.87		& 0.004 	&172		& -2.935	&&							& G		& 8867.5		& 0.006 	& 252	& -24.71	\\
July 18	& 						& G		& 8755.33		& 0.007 	& 231 	& -4.922	&&							& G		& 8868.03		& 0.001	& 141 	& -2.992	\\
\hline
\hline
\multicolumn{14}{l}{Table Comments: Measurement errors are $\pm$10\%. $^a$Lorentzian (L), Gaussian (G), complex (c) \bf $^b$Wavelengths are in observer frame.}\\
\end{tabular}
\label{table:line_IDs}
\end{table*}

\section{X-ray Analysis}\label{section_xray_spectra}
	\subsection{Spectral Modeling}
X-ray emission from ULX-1 was detected with a signal-to-noise $\geq3$ in all 11 \Swift/XRT observations obtained since 2016. High-quality ULX spectra typically require multiple components \citep{Sutton+13,Middleton+15a}. \citet{Carpano+18} found that variations in the observed flux from ULX-1 in \XMM and {\it NuSTAR} could be interpreted in changes in absorbing column only. We therefore simultaneously fit all eleven 0.3-10 keV \Swift spectra using  \texttt{XSPEC} \citep{Arnaud96} v.12.9.1 with a single model (\texttt{tbabs*pcfabs*(diskbb+nthcomp)}); only the covering fraction was allowed to vary between each observation. The neutral absorbing column lower limit was set to the Galactic line-of-sight value \citep[4.09$\times10^{20}$ cm$^{-2}$; ][]{Kalberla+05}. We used a multicolor disk blackbody component to model the optically thick wind emission, and \texttt{nthcomp} has been used previously to describe the harder X-ray emission originating in the inner disk \citep[e.g., ][while other authors have used a simple power law to represent this component]{Middleton+15a}. Due to the low number of counts, we fit the unbinned spectra using $C$-statistics, and the Pearson \chisq statistic was used as a ``test'' statistic for comparison purposes only.

The best-fit spectral model parameters are given in Table~\ref{table:best_spectral_fit}. The fit parameters were generally consistent with those found by \citet{Carpano+18}, although the \Swift data require substantial intrinsic absorption associated with the partial covering model component ($>32.2\times10^{22}$ cm$^{-2}$). A similarly high absorbing column was required to adequately fit the 2010 XMM observations that were obtained shortly after the initial outburst. This result is consistent with the recent multiwavelength SED modeling by \citet{Lau+16} and \citet{Villar+16}, which suggested that dust was actively re-forming in the vicinity of ULX-1. Details about the eleven individual observations used in the spectral fitting, with the observation-specific best-fit partial covering fraction and unabsorbed luminosities, are presented in Table~\ref{table:spectral_fits}.

\begin{figure}
\centering
\includegraphics[width=1\linewidth,clip=true,trim=0cm 0cm 0cm 0cm]{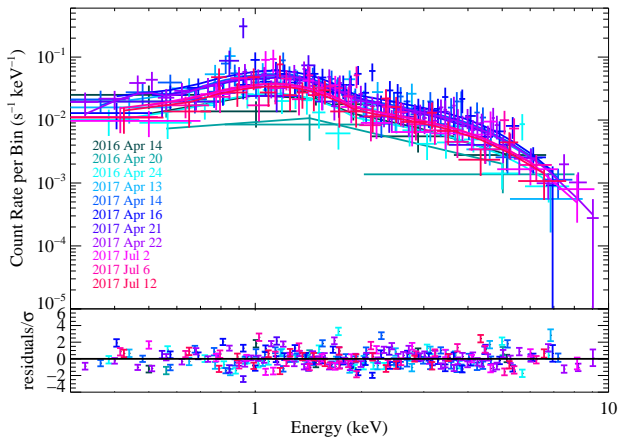} 
\caption{The 0.3-10 keV \Swift/XRT spectra, with the best-fit model superimposed. Spectra are color-coded by observation date.}
\label{figure:spectral_fits}
\end{figure}

Figure~\ref{figure:spectral_fits} shows the 0.3-10 keV spectra for all eleven observations, binned for display purposes only, with the best-fit models superimposed. The unabsorbed luminosities are essentially constant (within a factor of $\sim$3) across all observations. The average unabsorbed 0.3-10 keV luminosity is $(3.5\pm1.0)\times10^{39}$ \lum, similar to what was measured by \citet{Carpano+18} and roughly an order of magnitude higher than the classical Eddinginton limit for a neutron star. We can use the luminosity relation $L_X=GM_{\rm NS}\dot{m}R_{\rm NS}^{-1}$ to estimate the average mass accretion rate onto the ULX-1 neutron star. Assuming canonical neutron star values of $M_{\rm NS}=1.4$ \Msun and $R_{\rm NS}$=10 km, we estimate a mass accretion rate $\dot{m} \sim3\times10^{-7}$ \Msun yr$^{-1}$.

\begin{figure}
\centering
\includegraphics[width=1\linewidth,clip=true,trim=0cm 0cm 0cm 0cm]{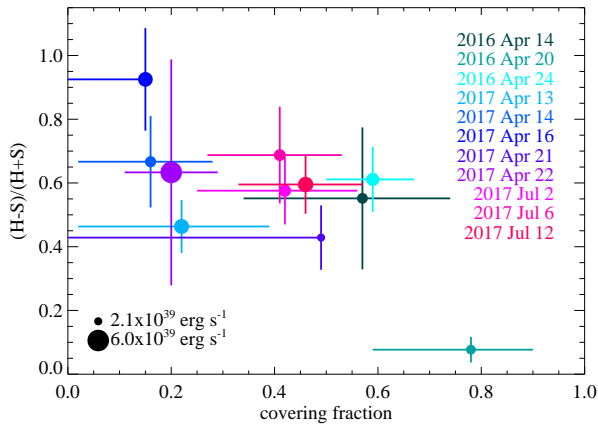} 
\caption{The spectral hardness ratio $HR$ as a function of covering fraction. The symbols are scaled to represent the 0.3-10 keV flux, and color-coded according to observation date.}
\label{figure:hr_cvrfract}
\end{figure}

The partial covering fractions vary significantly between observations, from $\sim$78\% on 2016 April 04 to consistent with $\sim$0\% a year later. We define the spectral hardness ratio $HR = (H-S)/(H+S)$, where $H$ is the unabsorbed flux in the 1-10 keV band and $S$ is the unabsorbed flux in the 0.3-1 keV band. There is evidence for changes in the unabsorbed spectral hardness with partial covering fraction (see Figure~\ref{figure:hr_cvrfract}) -- the observation with the highest covering fraction exhibits the lowest value of $HR$ (e.g., the unabsorbed hard flux was comparable to the unabsorbed soft flux), while observations with smaller covering fractions exhibit significantly harder values of $HR$.

\begin{table}
\centering
\caption{Best-Fit 0.3-10 keV Spectral Model}
\begin{tabular}{cccccccccccccc}
\hline \hline
Component	& Parameter		& Best-Fit Value	\\
(1)			& (2)				& (3)				\\
\hline
\texttt{tbabs}		& \nH (10$^{22}$ cm$^{-2}$)	& 0.16$\pm$0.03		\\
\texttt{pcfabs}		& \nH (10$^{22}$ cm$^{-2}$)	& $>$32.2				\\
\texttt{diskbb}		& $kT_{\rm in}$ (keV)		& 0.40$^{+0.08}_{-0.06}$	\\
\texttt{nthcomp}		& $\Gamma$				& 1.54$^{+0.11}_{-0.09}$	\\
\texttt{nthcomp}		& $kT_e$ (keV)				& 1.76$^{+2.53}_{-0.41}$	\\
\hline
				& 0.3-10 keV (\lum)			& $(3.5\pm1.0)\times10^{39}$	\\
Luminosity		& 0.3-1 keV (\lum)			& $(7.5\pm3.0)\times10^{38}$	\\
				& 1-10 keV (\lum)			& $(2.8\pm1.0)\times10^{39}$	\\
\hline
Photon Flux		& 54-200 eV (ph s$^{-1}$)		& $(7.2\pm1.3)\times10^{49}$	\\
\hline
				& degrees of freedom		& 1284				\\
Fit statistics		& $C$					& 1053.8				\\
				& $\chi^2_r$				& 1.00				\\
\hline \hline
\end{tabular}
\label{table:best_spectral_fit}
\end{table}

\begin{table*}\setlength{\tabcolsep}{2.2pt}
\scriptsize
\centering
\caption{\Swift/XRT Observation Log, Covering Fractions, and Variability}
\begin{tabular}{cccccccccccccc}
\hline \hline
Obs. ID		& Date	& Exp. 		&  \multicolumn{2}{c}{Position (J2000)}	&& Net Counts		& covering	& $F_{\rm var}$		& \multicolumn{3}{c}{Luminosity (10$^{39}$ \lum)}	&& $HR$		\\ \cline{4-5} \cline{10-12}
			&		& Time (s) 	& R.A. 			& Decl. 			&& (0.3-10 keV)	& fraction (\%)	& (\%)			& 0.3-10 keV	& 0.3-1 keV	& 1-10 keV		&&			\\
(1)			& (2)		& (3)			& (4)				& (5)				&& (6)			& (7)			& (8)				& (9)			& (10)		& (11)			&& (12)		\\
\hline

00049834002	& 2016-04-14	& 599	& 00:55:04.9	& -37:41:40.2	&& 43$\pm$9		 & 57$^{+17}_{-23}$	& 33$\pm$11	& $2.9^{+16.2}_{-2.4}$	& $0.6^{+0.4}_{-0.3}$	& $2.2^{+1.1}_{-0.8}$	&& 0.55$\pm$0.22	\\ 
00049834003	& 2016-04-20	& 489	& 00:55:04.7	& -37:41:40.3	&& 156$\pm$5		 & 78$^{+12}_{-19}$	& 31$\pm$8	& $2.6^{+14.4}_{-2.2}$	& $1.2^{+0.9}_{-0.6}$	& $1.4^{+1.9}_{-0.8}$	&& 0.08$\pm$0.04	\\
00049834005	& 2016-04-24	& 2922	& 00:55:04.9	& -37:41:46.1	&& 194$\pm$19	 & 59$^{+8}_{-9}$	& 33$\pm$11	& $3.6\pm0.5$ 			& $0.7\pm0.2$			& $2.9\pm0.5$			&& 0.61$\pm$0.10	\\
00049834006	& 2017-04-13	& 1128	& 00:55:04.9	& -37:41:41.0	&& 119$\pm$14	 & 22$^{17}_{-20}$	& 32$\pm$11	& $4.1\pm0.7$			& $1.1\pm0.3$			& $3.0^{+0.8}_{-0.6}$	&& 0.46$\pm$0.08	\\
00049834007	& 2017-04-14	& 2545	& 00:55:04.8	& -37:41:46.7	&& 338$\pm$25	 & 16$^{+12}_{-14}$	& 29$\pm$12	& $3.0^{+0.4}_{-0.3}$	& $0.5\pm0.2$			& $2.5^{+0.4}_{-0.3}$	&& 0.67$\pm$0.14	\\
00049834008	& 2017-04-16	& 1853	& 00:55:04.5	& -37:41:48.6	&& 301$\pm$23	 & $<20$			& 34$\pm$9	& $4.0^{+0.5}_{-0.4}$	& $0.25\pm0.03$		& $4.0^{_0.5}_{-0.3}$	&& 0.93$\pm$0.16	\\
00049834009	& 2017-04-21	& 522	& 00:55:04.8	& -37:41:48.6	&& 47$\pm$9		 & $<49$			& 31$\pm$10	& $2.1^{+0.7}_{-0.6}$	& $0.6^{+0.3}_{-0.2}$	& $1.5^{+0.6}_{-0.5}$	&& 0.43$\pm$0.10	\\
00049834010	& 2017-04-22	& 5175	& 00:55:04.5	& -37:41:46.6	&& 617$\pm$33	 & $20\pm9$		& 31$\pm$10	& $6.0^{+7.2}_{-3.2}$	& $1.1^{+1.5}_{-0.6}$	& $4.9^{+5.6}_{-2.6}$	&& 0.63$\pm$0.35	\\
00049834012	& 2017-07-02	& 1296	& 00:55:04.7	& -37:41:41.2	&& 116$\pm$14	 & 42$^{+14}_{-17}$	& 31$\pm$12	& $3.3^{+0.7}_{-0.6}$	& $0.7^{+0.3}_{-0.2}$	& $2.6^{+0.7}_{-0.6}$	&& 0.58$\pm$0.11	\\
00049834013	& 2017-07-06	& 1953	& 00:55:04.4	& -37:41:47.0	&& 187$\pm$18	 & 41$^{+12}_{-14}$	& 35$\pm$11	& $3.2^{+0.6}_{-0.5}$	& $0.5\pm0.2$			& $2.7\pm0.5$			&& 0.69$\pm$0.15	\\
00049834014	& 2017-07-12	& 1833	& 00:55:04.4	& -37:41:44.0	&& 156$\pm$17	 & 46$^{+11}_{-13}$	& 32$\pm$9	& $4.2^{+0.7}_{-0.6}$	& $0.8^{+0.3}_{-0.2}$	& $3.3^{+0.7}_{-0.6}$	&& 0.60$\pm$0.09	\\
\hline \hline
\end{tabular}
\label{table:spectral_fits}
\end{table*}

	\subsection{Timing}
X-ray variability provides additional clues to the geometry of supercritical accretion onto the neutron star. Variability can arise from clumps or inhomogeneities in the winds as they transit the central region from the perspective of the observer. Alternatively, inwardly propagating variations in surface density or mass accretion rate \citep[``flicker noise,'' e.g., as mass is lost to the wind,][]{Lyubarskii97,Ingram+12,Middleton+15a} can produce variations in X-ray emission, especially at sub-Eddington accretion rates \citep{Remillard+06,Heil+15a,Heil+15b}. Following the approach of \citet{Middleton+15a}, we compute the fractional root-mean-square variability ($F_{\rm var}$) by Fourier-transforming $\sim$1000 s segments of the background-subtracted 0.3-10 keV light curves of ULX-1 and averaging the resulting periodograms \citep[see also][]{vanderKlis89}. $F_{\rm var}$ is then calculated by integrating the power over the frequency range of 3--200 mHz \citep[where most broad-band variability is observed in ULXs][]{Heil+09} and taking the square root \citep{Edelson+02}. In all eleven \Swift observations, $F_{\rm var}\sim30\%$ (see Table~\ref{table:spectral_fits}). Similar variability behavior has been observed in other ULXs with relatively soft spectra \citep{Sutton+13,Middleton+15a}. We used a two-sided K-S test of the photon arrival time in each exposure against the assumption of a constant count rate, but do not find any evidence for significant variability on short time scales. Our light curves (shown in Figure~\ref{figure:timing_analysis}) are consistent with a constant count rate, with variations about a constant rate on the order of $\sim$30\%.

\begin{figure*}
\centering
\includegraphics[width=0.45\linewidth,clip=true,trim=0cm 0cm 0cm 0cm]{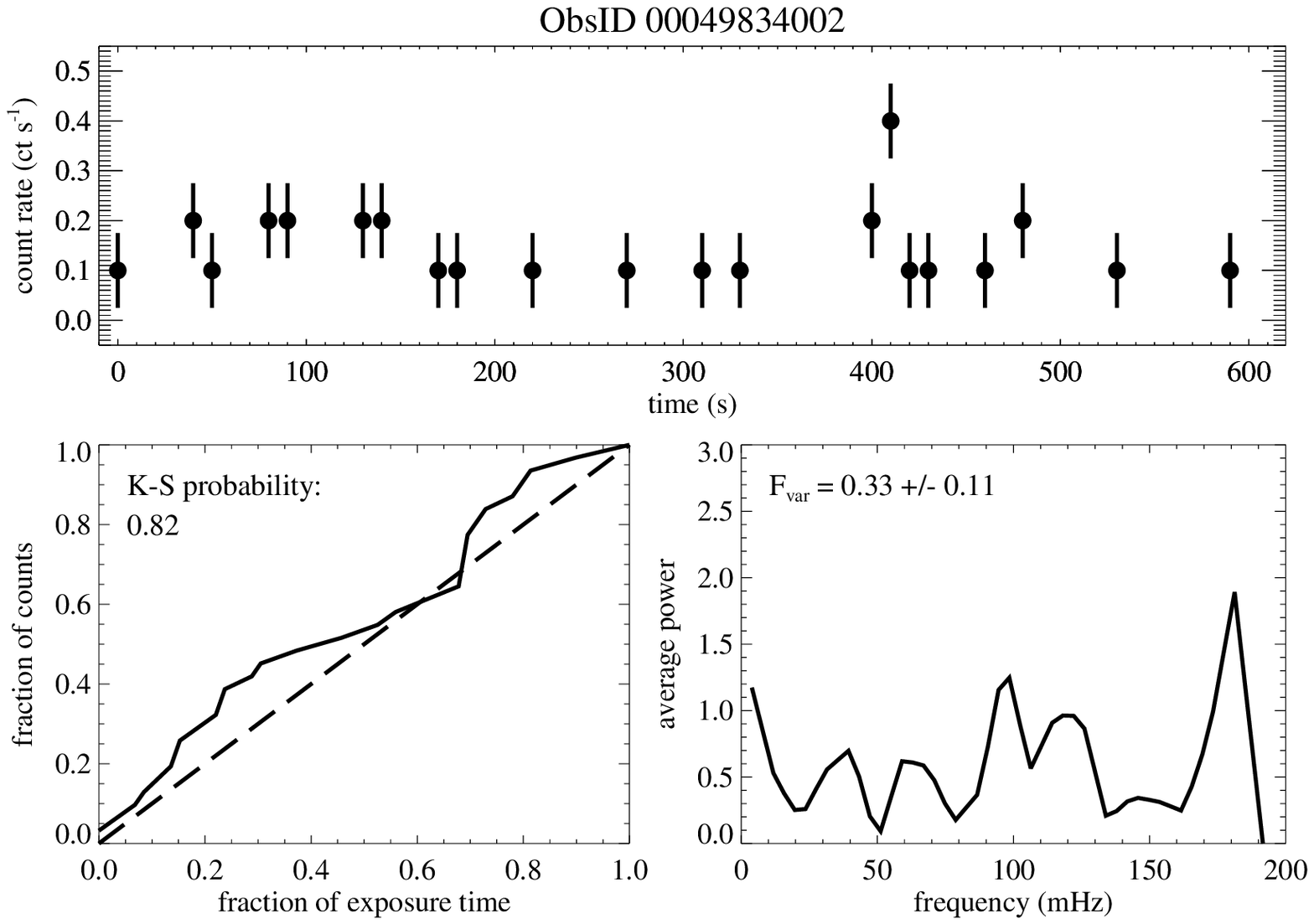} 
\includegraphics[width=0.45\linewidth,clip=true,trim=0cm 0cm 0cm 0cm]{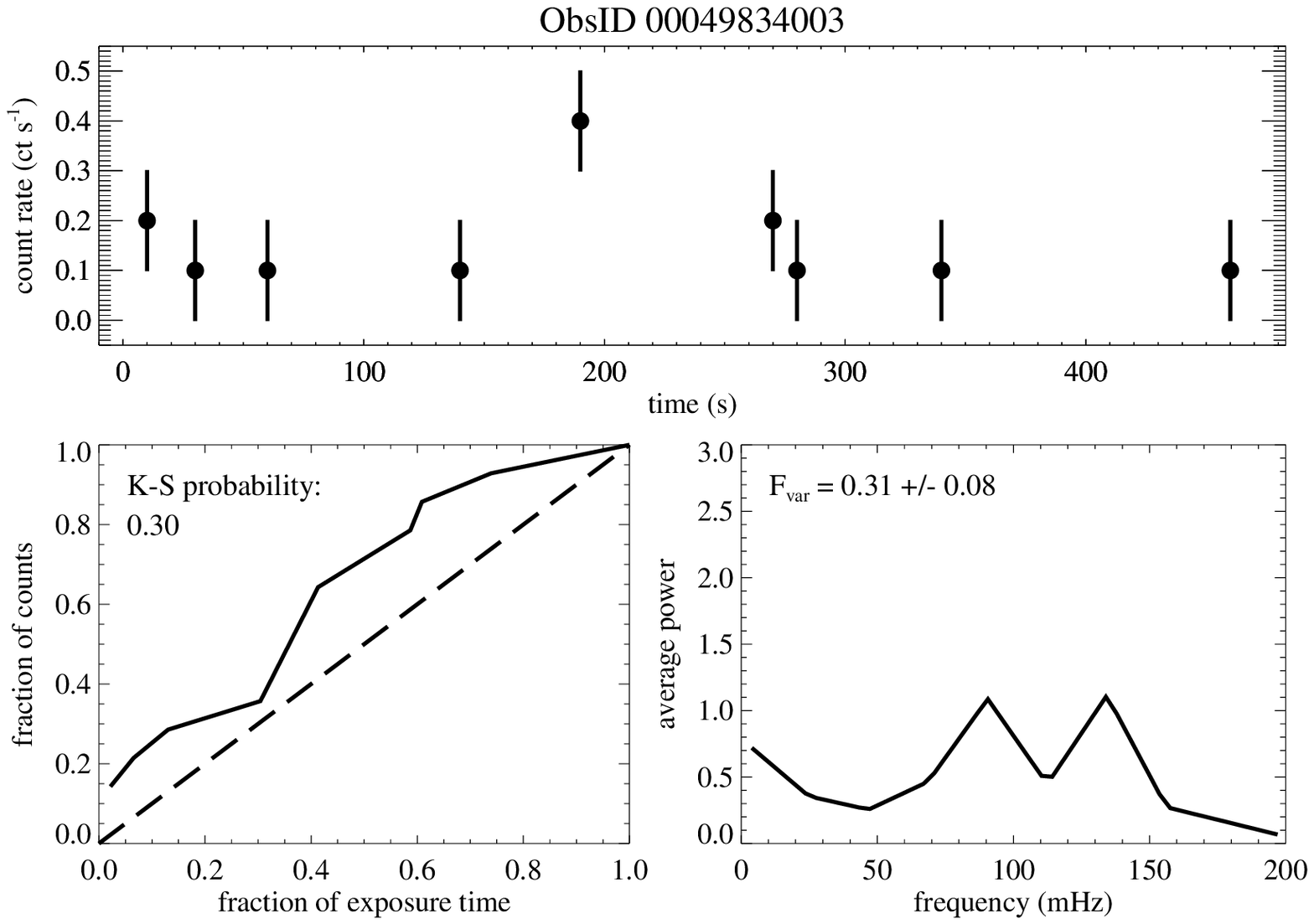} 

\includegraphics[width=0.45\linewidth,clip=true,trim=0cm 0cm 0cm 0cm]{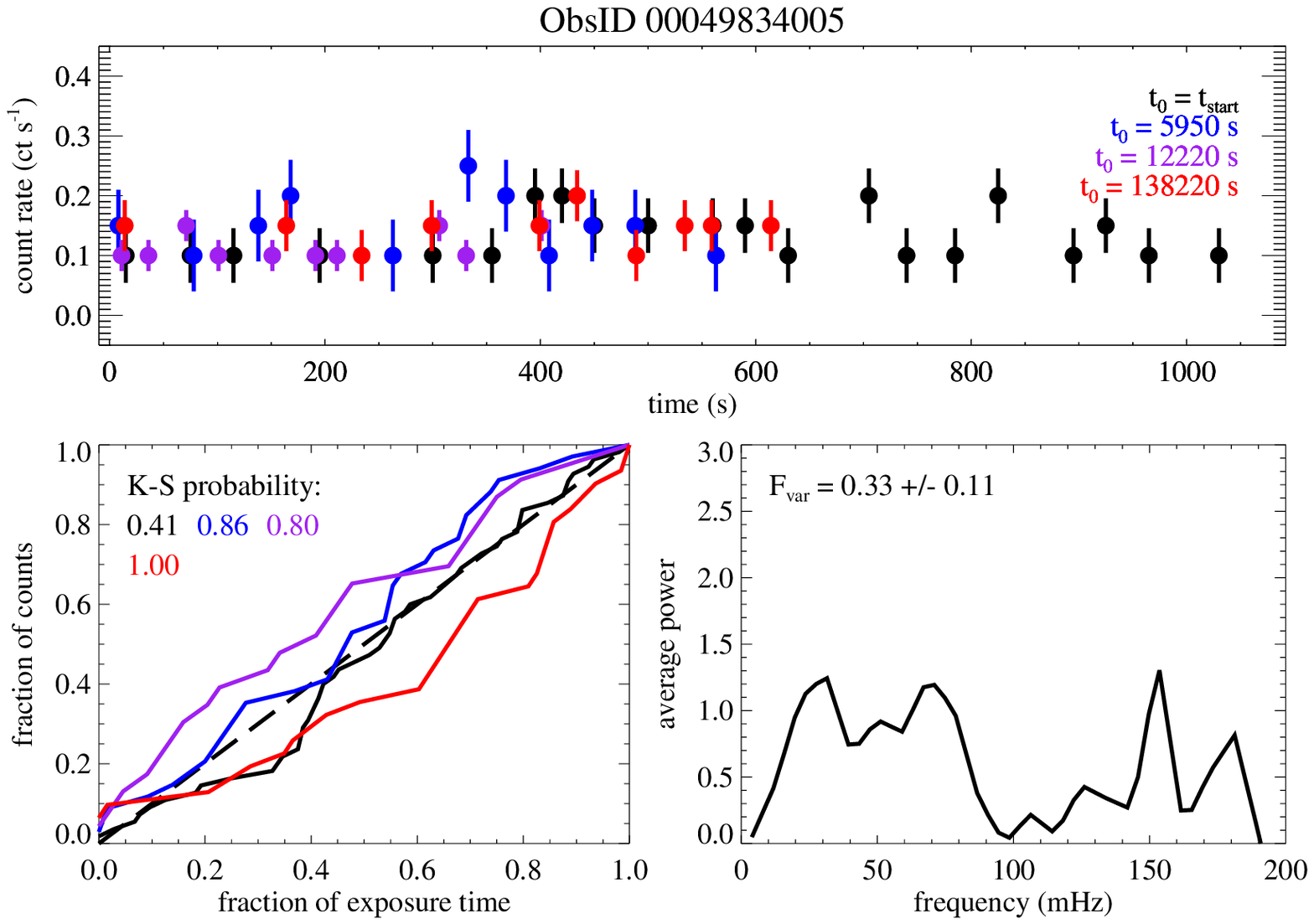} 
\includegraphics[width=0.45\linewidth,clip=true,trim=0cm 0cm 0cm 0cm]{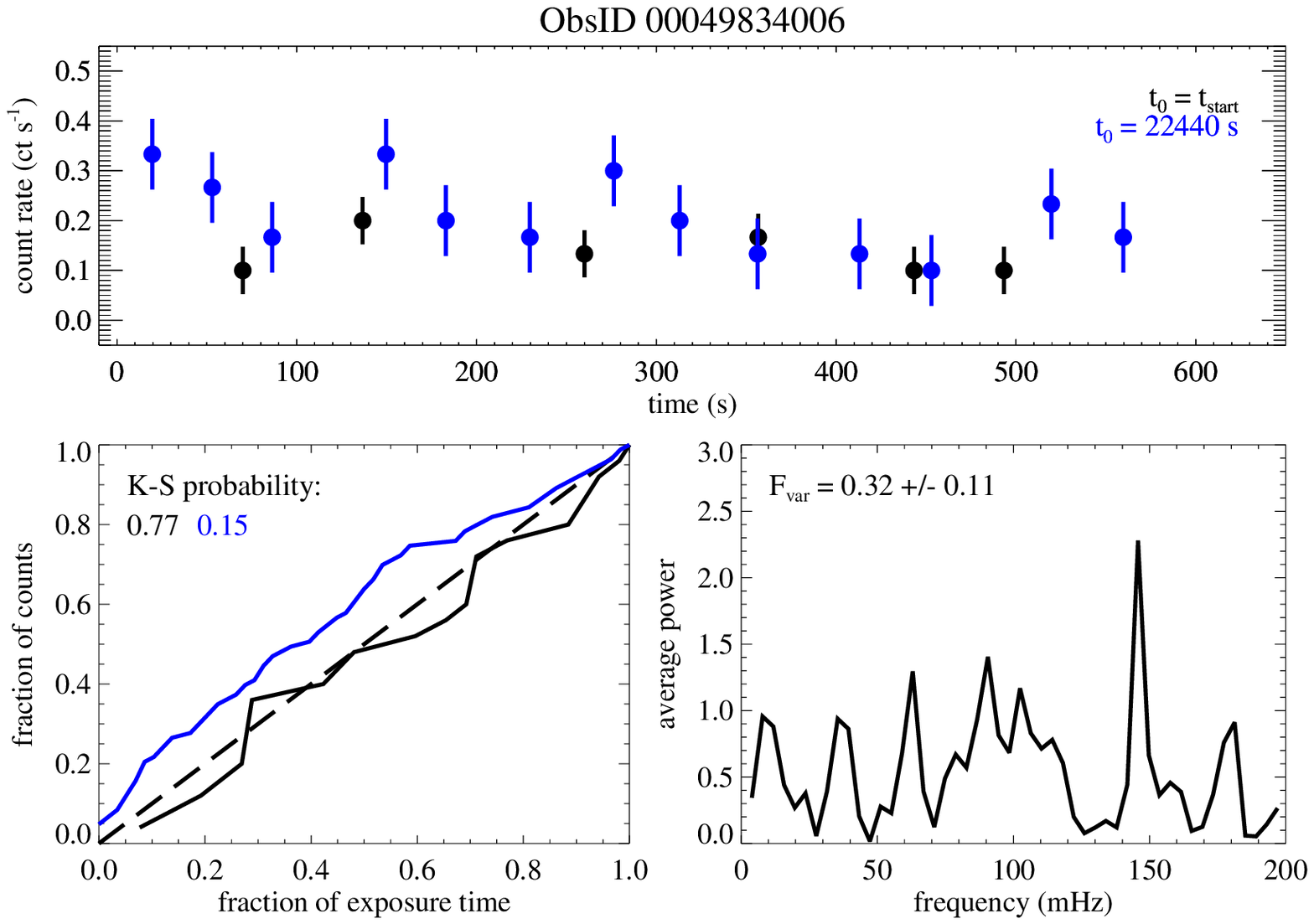} 

\includegraphics[width=0.45\linewidth,clip=true,trim=0cm 0cm 0cm 0cm]{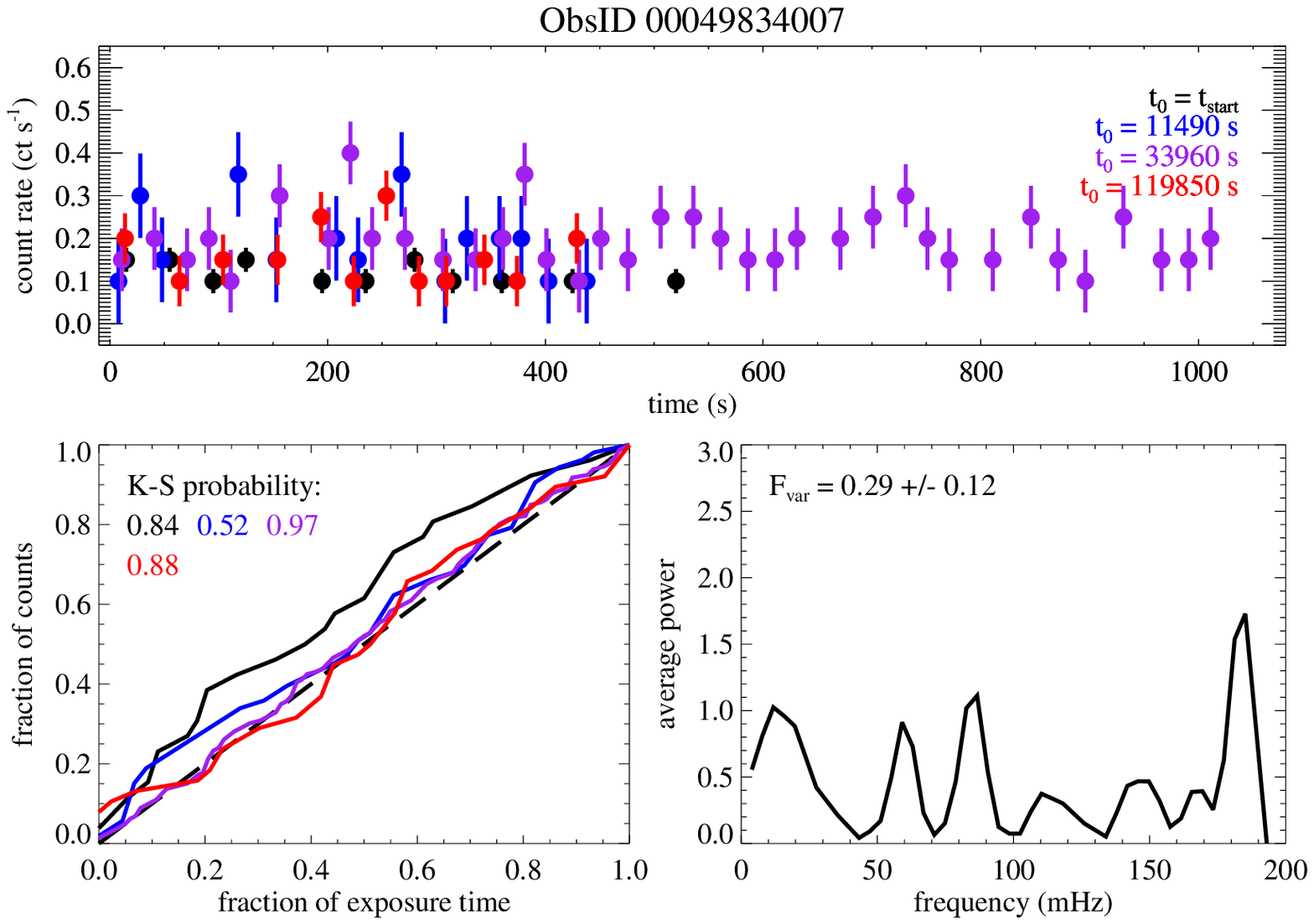} 
\includegraphics[width=0.45\linewidth,clip=true,trim=0cm 0cm 0cm 0cm]{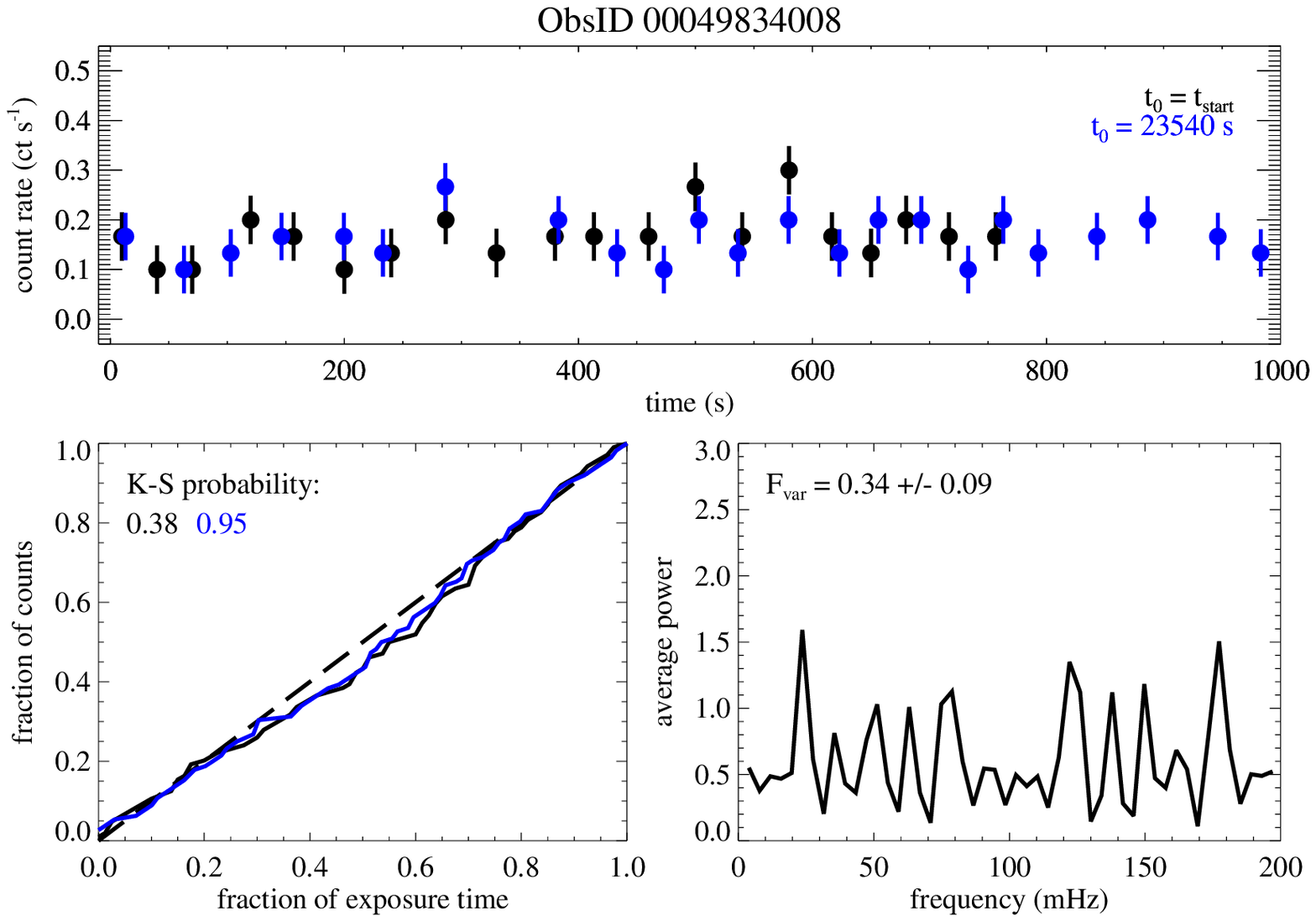} 
\caption{Timing analysis plots for the \Swift/XRT observations. The top panel shows the 0.3-10 keV light curve. The bottom-left panel shows the photon arrival time (solid line) versus the expectation for a constant count rate (dashed line), used to calculate the K-S probability of non-variability. The bottom right panel shows the 3-200 mHz periodogram, used to calculate $F_{\rm var}$ (see text for details).}
\label{figure:timing_analysis}
\end{figure*}

\begin{figure*}\setcounter{figure}{4}
\centering
\includegraphics[width=0.45\linewidth,clip=true,trim=0cm 0cm 0cm 0cm]{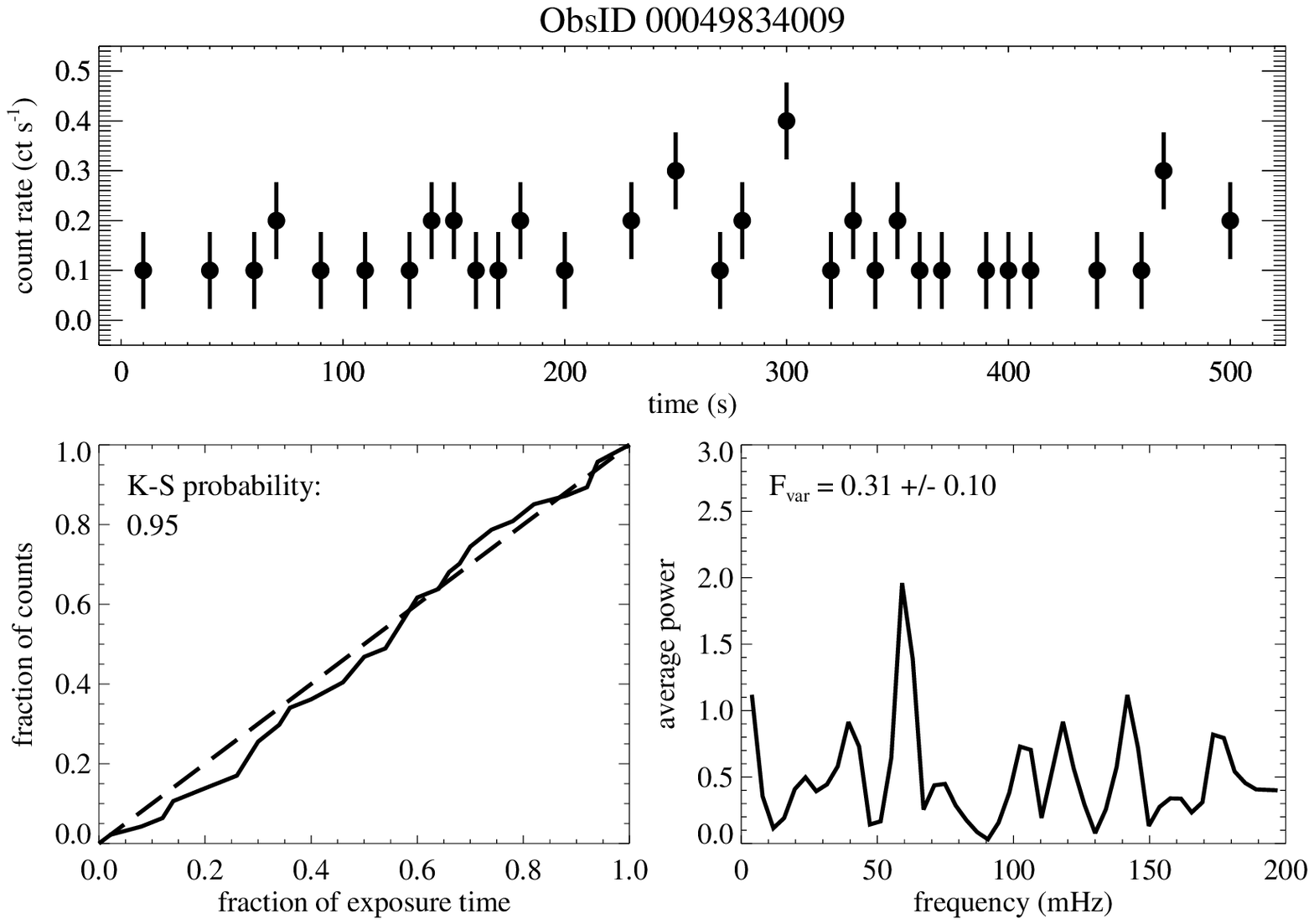} 
\includegraphics[width=0.45\linewidth,clip=true,trim=0cm 0cm 0cm 0cm]{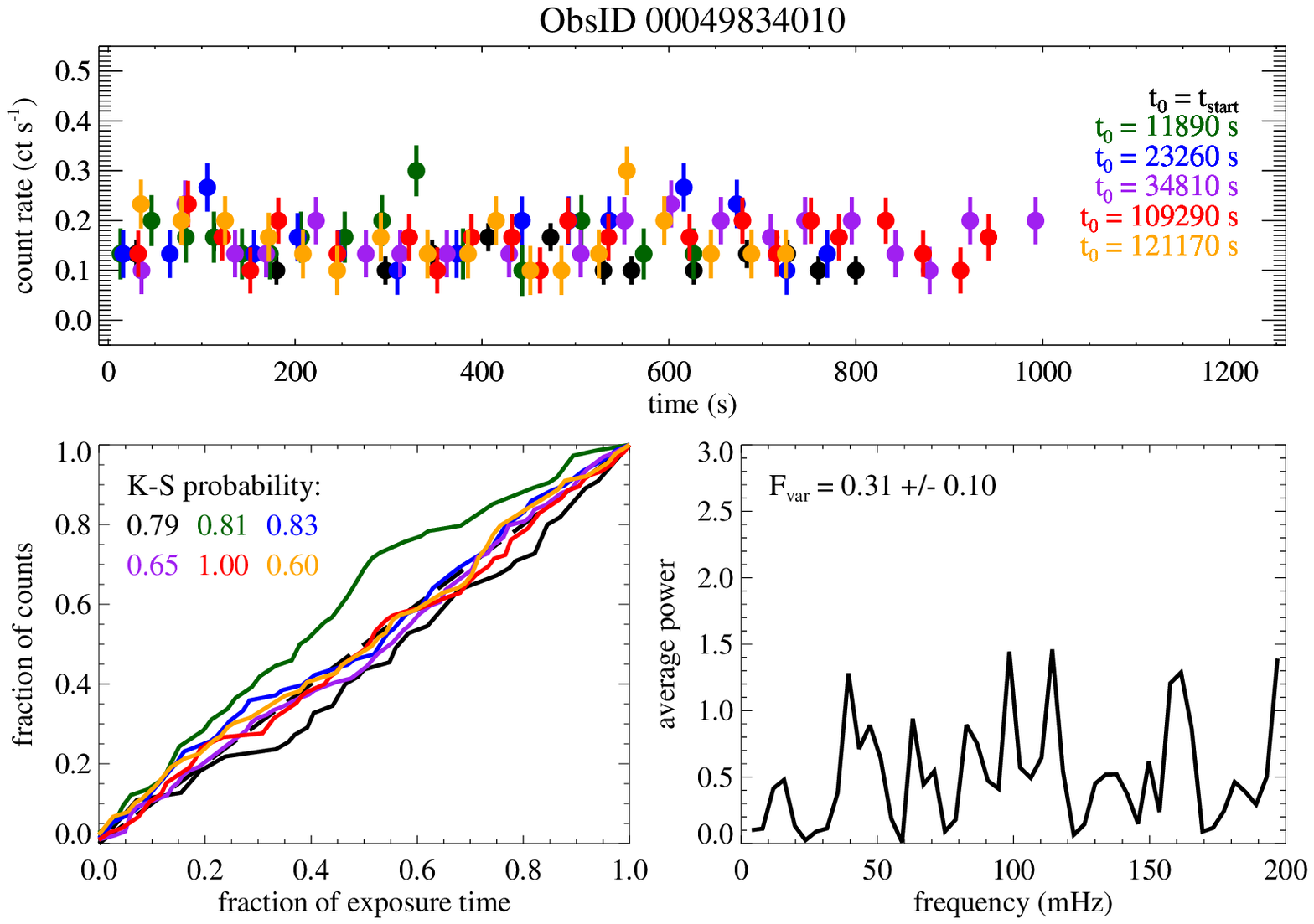} 

\includegraphics[width=0.45\linewidth,clip=true,trim=0cm 0cm 0cm 0cm]{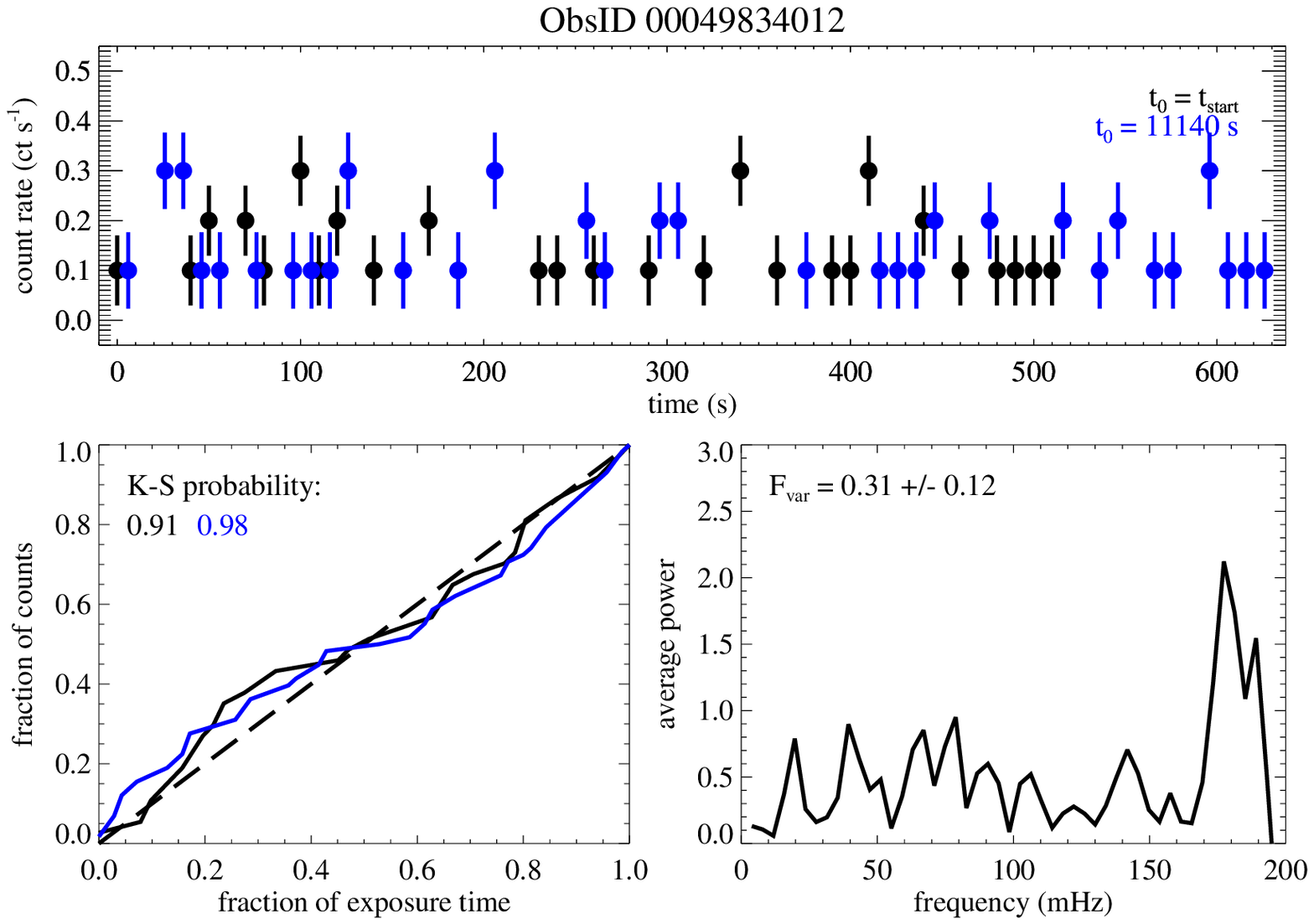} 
\includegraphics[width=0.45\linewidth,clip=true,trim=0cm 0cm 0cm 0cm]{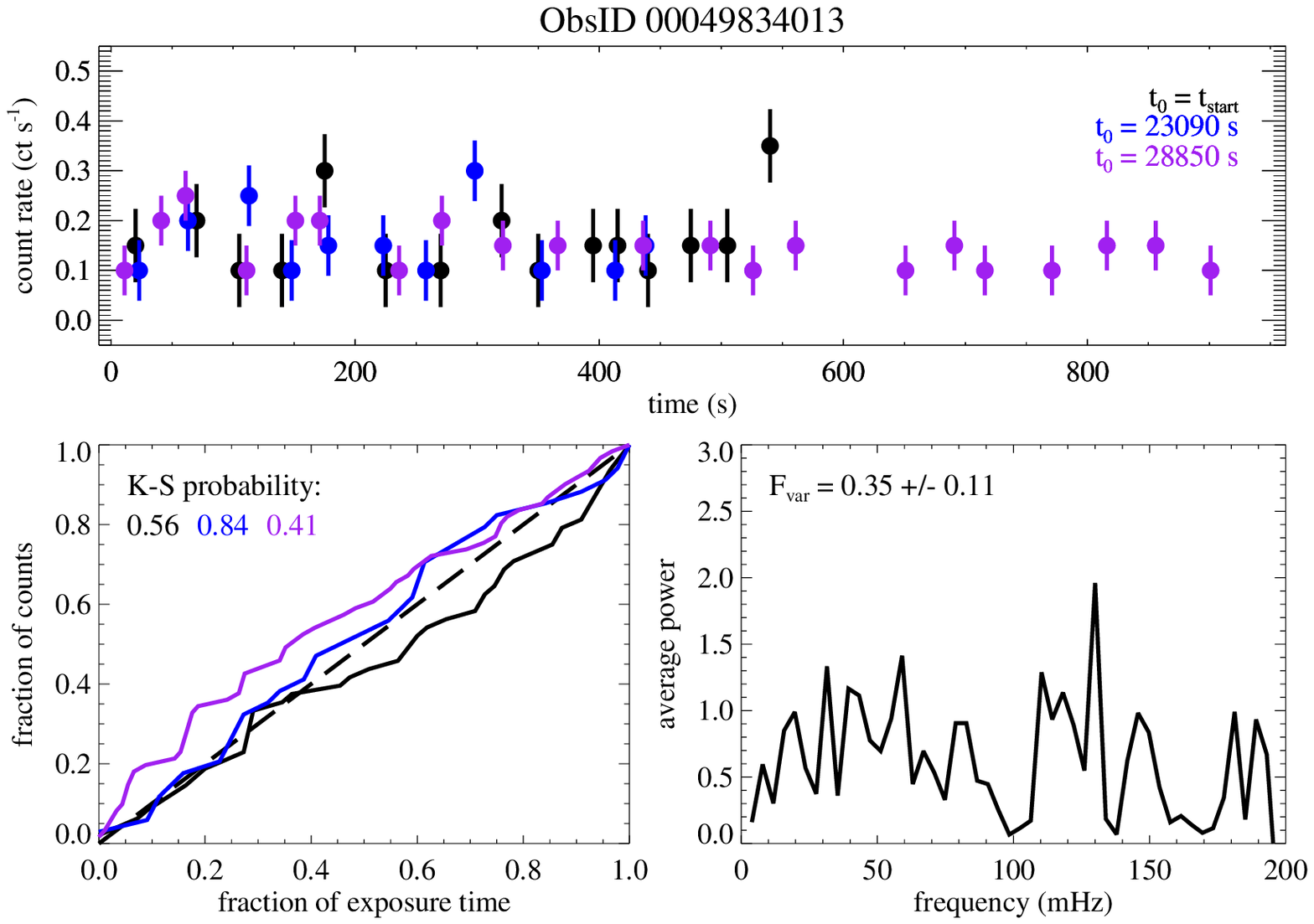} 

\includegraphics[width=0.45\linewidth,clip=true,trim=0cm 0cm 0cm 0cm]{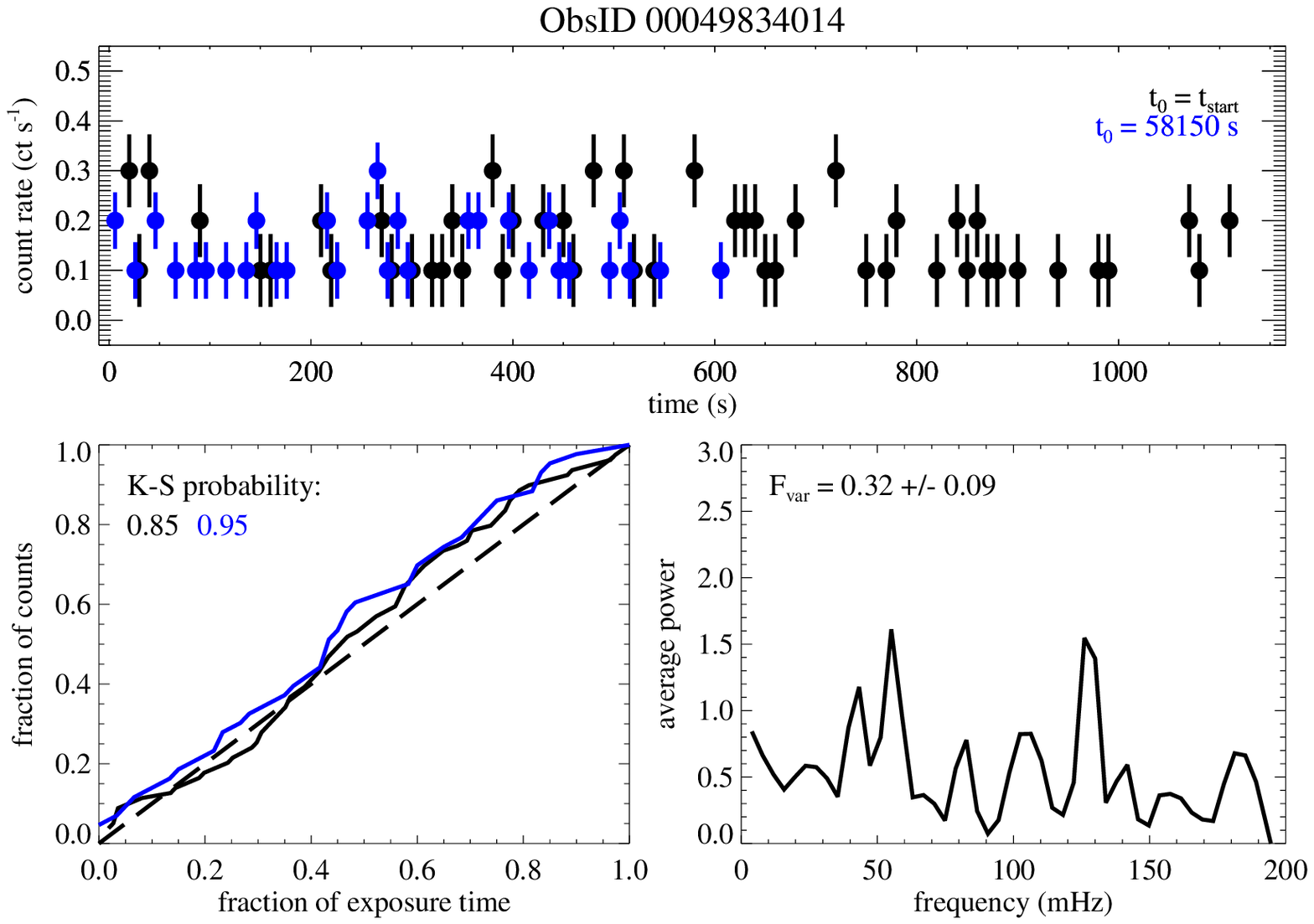} 
\caption{Timing analysis plots for the \Swift/XRT observations, continued.}
\label{figure:timing_analysis}
\end{figure*}

\section{Discussion}\label{section_discussion}
	\subsection{Potential Evidence for Geometric Beaming?}
The observed variation in covering fractions and spectral hardnesses from our X-ray spectral modeling is likely driven by changes in the mass accretion rate onto the ULX-1 neutron star primary. Using the spectral-timing model of \citet{Sutton+13} and \citet{Middleton+15a}, we can interpret the partial covering component of the spectral model as representing optically thick clumps in the winds of ULX-1 as they obscure the central ionizing source. High covering fractions indicate an increased homogeneity of the winds, and the simultaneous softening of the spectrum may correspond to an increased mass accretion rate. The high luminosities of ULXs have led some to speculate that beaming effects may be important for understanding the geometry of the underlying X-ray binary \citep{King09}; such beaming would occur if the observer's line of sight to the ULX ($\theta$) falls within the opening angle over which the wind is launched \citep[$\theta_{\rm w}; $i.e., see Figure~1 in][]{Middleton+15a}. At low inclinations, the observer will have a relatively unobstructed view of the hot inner accretion disk and will therefore observe a harder spectrum. Although there is evidence for a relatively hard component in the ULX-1 spectrum ($\Gamma\sim1.5$), $>$90\% of the unabsorbed flux originates in the soft wind component in all eleven observations. Our X-ray spectral-timing suggest that we are observing ULX-1 at small- to moderate-$\theta$.

The importance of beaming was explored for the pulsed neutron star-ULX M82~ULX-2 \citep{King+16}. The relationship between the observed X-ray luminosity and $\dot{m_0}$ (the ratio of the thermal timescale mass transfer rate to the Eddington rate) is given by \citep{King09,King+16}: 

\begin{equation}
\frac{m_1}{L_{40}} \approx \frac{4500}{\dot{m_0}^2 \left( 1+\text{ln} ~ \dot{m_0} \right),}
\end{equation}

\noindent where $m_1$ is the mass of the neutron star (assumed to be 1.4 \Msun) and $L_{40}$ is the observed X-ray luminosity in units of 10$^{40}$ \lum. Using the average X-ray luminosity from our spectral modeling ($L_{40}$ = 0.35) yields $\dot{m_0}\sim17$. \citet{King09} give an approximate relationship between $\dot{m_0}$ and the beaming factor $b$,

\begin{equation}
b\approx\frac{73}{\dot{m_0}^2},
\end{equation}

\noindent which is valid for $\dot{m_0}\gtrsim8.5$. We therefore estimate a modest beaming factor $b\sim0.25$ for ULX-1. The ``spherization'' radius $R_{\rm sph}$ from which the winds are launched can be estimated from \citet{Shakura+73}:

\begin{equation}
R_{\rm sph}\approx\frac{27}{4}\dot{m_0}R_g,
\end{equation}

\noindent where $R_g$ is the gravitational radius of the neutron star, $R_g=GM/c^2$. Using this expression, we estimate $R_{\rm sph}\sim240$ km.

Geometric beaming, if present, would further impart evidence on the surrounding material. ULXs are expected to ionize the surrounding ISM \citep{Tarter+69,Kallman+McCray82,Pakull+Angebault86} in a manner that is analogous to UV-ionized H~II regions. X-ray ionized nebula, however, will lack a clearly defined boundary \citep[e.g., the equivalent of a Str\"{o}mgren sphere; ][]{Pakull+Mirioni02} and will instead produce an extended, warm region of weakly ionized atoms and collisionally-excited species. The resulting X-ray-ionized nebula will exhibit spectral features from high-ionization species, notably the He~II $\lambda$4686 emission line.

The He~II $\lambda$4686 emission line is particularly sensitive to the photon flux originating from 54-200 eV. We measure an average FWHM of $\sim$345 km s$^{-1}$ in the three Gemini spectra where the He~II $\lambda$4686 line is measurable, somewhat higher than the $\sim$270 km s$^{-1}$ FWHM value found by \citet{Villar+16}. Following the approach of \cite{Pakull+Angebault86}, we can combine our \Swift X-ray observations and Gemini optical spectroscopy to determine to what degree beaming is occurring in the ULX-1 system. The 54--200 eV photon flux ($Q$) can be directly measured from our X-ray spectra. These X-ray photons will then excite the He~II $\lambda$4686 emission line in the Gemini spectrum. The absorbed rate of He$^+$ Lyman continuum photons, $Q'$, in the nebula surrounding ULX-1 can then be calculated by

\begin{equation}
Q^{\prime} = \frac{L_{\rm 4686}}{h\nu_{\rm 4686}} \frac{\alpha_{\rm B}\left(He^{+}, T \right)}{\alpha^{\rm eff}_{\rm 4686}\left( T \right)},
\end{equation}

\noindent where $\alpha_{\rm B}$(He$^{+}, T$) is the recombination coefficient summed over all levels above the ground state and $\alpha^{\rm eff}_{\rm 4686} (T)$ is the effective recombination coefficient for the emission of He~II $\lambda$4686 photons, which carry an energy $h\nu_{\rm 4686}$. The ratio of these two coefficients is $\sim$5.2 \citep{Pakull+Angebault86}, and depends only weakly on the electron temperature.

The observed He~II luminosity from the optical spectrum obtained on 2 July 2017 is $L_{\rm 4686}\sim7-9\times10^{35}$ \lum, which implies $Q^{\prime}\sim1.1\times10^{48}$ ph s$^{-1}$. The \Swift X-ray spectrum obtained on the same day yields $Q\sim$6.4$\times10^{47}$ ph s$^{-1}$; in general, the X-ray spectra predict an average $Q$ of $(6.8\pm2.8)\times10^{47}$ ph s$^{-1}$. The X-ray predicted $Q$ is therefore broadly consistent with the observed $Q^{\prime}$. Any geometric beaming effects, if present in the ULX-1 system at all, are likely small. 
 
 	\subsection{The Origin of the Optical Emission Lines}
Numerous emission lines are present in the Gemini spectra. Optical images of ULX-1 obtained in 2016 with the {\it Hubble Space Telescope} did not show evidence for significant nebulosity -- the extinction intrinsic to NGC~300 at the location of the X-ray source was estimated to be $A_V\sim0.4$ mag \citep{Binder+16}. Recent work by \citet{Lau+16} and \citet{Villar+16} indicates that dust may be actively reforming in the vicinity of the central binary (within $\sim$few hundred AU). We therefore expect the emission lines observed in the Gemini spectra originate from the ULX-1 binary or its immediate vicinity, and not from an extended nebula.

The lack of strong [O II] and [S II] lines imply that the material in the immediate vicinity of ULX-1 is not shock-heated \citep{Baldwin+81,Sutherland+93,Kewley+01}. The [O III] $\lambda$5007/H$\beta\sim0.5$ line ratio and lack of detectable and [N II] $\lambda$6583 places ULX-1 firmly the photoionization-dominant quadrant of the classic BPT diagram \citep[see, e.g.,][and references therein]{Evans+99,Abolmasov+07}. This differs from several ULXs that are known to be associated with bright nebulae and bubbles, which typically exhibit spectral features consistent with bright, shock-powered nebulae \citep{Pakull+Mirioni03,Pakull+06}. Ionizing X-ray and UV radiation originating from the 2010 outburst could have affected the surrounding environment only out to a radius of $\sim$8 light years ($\sim$2.5 pc); at the distance of NGC~300, this distance corresponds to an angular size of 0\farcs26, less than the width of the slit used to obtain the Gemini spectra. The lack of a bright, shock-powered nebula may be due to the young age of the system: ULX-1 is known to be associated with a very young ($<$5 Myr) stellar population \citep{Binder+16}. Although this young age is not unique among ULXs \citep[several nearby ULXs also exhibit extremely young ages, consistent with X-ray binary and ULX formation models;][]{Poutanen+13,Berghea+13,Wiktorowicz+17,Linden+10}, most ULXs with extended nebulae are >10 Myr old \citep[see][and references therein]{Berghea+13,Voss+11,Yang+11} and have been presumably producing ionizing radiation over significantly longer timescales.

H$\alpha$ is well described by a Lorentzian profile with a full width at half maximum (FWHM) of $\sim$300 km s$^{-1}$ \citep[compared to $\sim$560 km s$^{-1}$ following the initial outburst;][]{Villar+16}. The H$\alpha$ flux varies by a factor of $\sim$4 across our observations, ranging from 7.8$\times10^{-15}$ \flux \AA$^{-1}$ to 3.1$\times10^{-14}$ \flux \AA$^{-1}$. In all four observations, the flux ratio  H$\alpha$/H$\beta>4$. This is similar to the late-time spectrum obtained by \citet{Villar+16}, and is consistent with emission from a persistent wind or mass loss from a blue or yellow supergiant donor star (and not standard case B recombination). A detailed study of the optical spectra will be presented in a forthcoming paper.

\section{Conclusions}
We have obtained new, near-simultaneous \Swift/XRT imaging and Gemini GMOS spectroscopy for the supernova impostor-turned-ULX SN~2010da in NGC~300. The X-ray emission from the object has been persistently high, $\sim2-6\times10^{39}$ \lum, since 2016, consistent with deep \XMM+{\it NuSTAR} observations \citep{Carpano+18}. The X-ray spectra and temporal properties are consistent with the widely accepted ``supercritical'' model of ULX accretion, in which optically thick winds are launched from a region close to the ionizing central source and reveal a hot, inner accretion disk. The optical spectra suggest that the neutron star X-ray source is photoionizing material in immediate vicinity ($<$2.5 pc) of the central binary.  Our comparison of the soft X-ray emission to the observed He~II $\lambda$4686 emission line luminosity suggests that geometric beaming effects are minimal in the ULX-1 system, making ULX-1 one of only a few bona fide ULXs to be powered by accretion onto a neutron star.

\section*{Acknowledgements}
This research has made by using \Swift/XRT data provided by the UK Swift Science Data Centre at the University of Leicester. The authors would like to thank Loredana Vetere for her assistance obtaining the \Swift ToO observations. We thank the \Swift team, the PI, the duty scientists and science planners for making the three ToO observations reported here possible. BB was supported by the NSF via award CAREER-1454333 (PI M. Povich).


\begin{thebibliography}{}

\bibitem[\protect\citeauthoryear{{Abolmasov} et~al.}{{Abolmasov}
  et~al.}{2007}]{Abolmasov+07}
{Abolmasov}, P., {Fabrika}, S., {Sholukhova}, O.,  \& {Afanasiev}, V. 2007,
  Astrophysical Bulletin, 62, 36

\bibitem[\protect\citeauthoryear{{Arnaud}}{{Arnaud}}{1996}]{Arnaud96}
{Arnaud}, K.~A. 1996, in Astronomical Society of the Pacific Conference Series,
  Vol. 101, Astronomical Data Analysis Software and Systems V, ed. G.~H.
  {Jacoby} \& J.~{Barnes}, 17

\bibitem[\protect\citeauthoryear{{Bachetti} et~al.}{{Bachetti}
  et~al.}{2014}]{Bachetti+14}
{Bachetti}, M., et~al. 2014, \nat, 514, 202

\bibitem[\protect\citeauthoryear{{Baldwin}, {Phillips}, \&
  {Terlevich}}{{Baldwin} et~al.}{1981}]{Baldwin+81}
{Baldwin}, J.~A., {Phillips}, M.~M.,  \& {Terlevich}, R. 1981, \pasp, 93, 5

\bibitem[\protect\citeauthoryear{{Baldwin} \& {Stone}}{{Baldwin} \&
  {Stone}}{1984}]{Baldwin+Stone84}
{Baldwin}, J.~A.,  \& {Stone}, R.~P.~S. 1984, \mnras, 206, 241

\bibitem[\protect\citeauthoryear{{Berghea} et~al.}{{Berghea}
  et~al.}{2013}]{Berghea+13}
{Berghea}, C.~T., {Dudik}, R.~P., {Tincher}, J.,  \& {Winter}, L.~M. 2013,
  \apj, 776, 100

\bibitem[\protect\citeauthoryear{{Binder} et~al.}{{Binder}
  et~al.}{2011}]{Binder+11}
{Binder}, B., {Williams}, B.~F., {Kong}, A.~K.~H., {Gaetz}, T.~J., {Plucinsky},
  P.~P., {Dalcanton}, J.~J.,  \& {Weisz}, D.~R. 2011, \apjl, 739, L51

\bibitem[\protect\citeauthoryear{{Binder} et~al.}{{Binder}
  et~al.}{2016}]{Binder+16}
{Binder}, B., {Williams}, B.~F., {Kong}, A.~K.~H., {Gaetz}, T.~J., {Plucinsky},
  P.~P., {Skillman}, E.~D.,  \& {Dolphin}, A. 2016, \mnras, 457, 1636

\bibitem[\protect\citeauthoryear{{Brown}}{{Brown}}{2010}]{Brown10}
{Brown}, P.~J. 2010, The Astronomer's Telegram, 2633, 1

\bibitem[\protect\citeauthoryear{{Burrows} et~al.}{{Burrows}
  et~al.}{2004}]{Burrows+04}
{Burrows}, D.~N., et~al. 2004, in \procspie, Vol. 5165, X-Ray and Gamma-Ray
  Instrumentation for Astronomy XIII, ed. K.~A. {Flanagan} \& O.~H.~W.
  {Siegmund}, 201

\bibitem[\protect\citeauthoryear{{Carpano} et~al.}{{Carpano}
  et~al.}{2018}]{Carpano+18}
{Carpano}, S., {Haberl}, F., {Maitra}, C.,  \& {Vasilopoulos}, G. 2018, \mnras

\bibitem[\protect\citeauthoryear{{Chornock} \& {Berger}}{{Chornock} \&
  {Berger}}{2010}]{Chornock+10}
{Chornock}, R.,  \& {Berger}, E. 2010, The Astronomer's Telegram, 2637, 1

\bibitem[\protect\citeauthoryear{{Chornock}, {Czekala}, \& {Berger}}{{Chornock}
  et~al.}{2011}]{Chornock+11}
{Chornock}, R., {Czekala}, I.,  \& {Berger}, E. 2011, The Astronomer's
  Telegram, 3726, 1

\bibitem[\protect\citeauthoryear{{Dalcanton} et~al.}{{Dalcanton}
  et~al.}{2009}]{Dalcanton+09}
{Dalcanton}, J.~J., et~al. 2009, \apjs, 183, 67

\bibitem[\protect\citeauthoryear{{Edelson} et~al.}{{Edelson}
  et~al.}{2002}]{Edelson+02}
{Edelson}, R., {Turner}, T.~J., {Pounds}, K., {Vaughan}, S., {Markowitz}, A.,
  {Marshall}, H., {Dobbie}, P.,  \& {Warwick}, R. 2002, \apj, 568, 610

\bibitem[\protect\citeauthoryear{{Elias-Rosa}, {Mauerhan}, \& {van
  Dyk}}{{Elias-Rosa} et~al.}{2010a}]{EliasRosa+10a}
{Elias-Rosa}, N., {Mauerhan}, J.~C.,  \& {van Dyk}, S.~D. 2010a, Central Bureau
  Electronic Telegrams, 2292, 2

\bibitem[\protect\citeauthoryear{{Elias-Rosa}, {Mauerhan}, \& {van
  Dyk}}{{Elias-Rosa} et~al.}{2010b}]{EliasRosa+10b}
{Elias-Rosa}, N., {Mauerhan}, J.~C.,  \& {van Dyk}, S.~D. 2010b, The
  Astronomer's Telegram, 2636, 1

\bibitem[\protect\citeauthoryear{{Evans} et~al.}{{Evans}
  et~al.}{1999}]{Evans+99}
{Evans}, I., {Koratkar}, A., {Allen}, M., {Dopita}, M.,  \& {Tsvetanov}, Z.
  1999, \apj, 521, 531

\bibitem[\protect\citeauthoryear{{Evans} et~al.}{{Evans}
  et~al.}{2009}]{Evans+09}
{Evans}, P.~A., {Beardmore}, A.~P., {Page}, K.~L., {Osborne}, J.~P., {Burrows},
  D.~N.,  \& {Gehrels}, N. 2009, in American Institute of Physics Conference
  Series, Vol. 1133, American Institute of Physics Conference Series, ed.
  C.~{Meegan}, C.~{Kouveliotou}, \& N.~{Gehrels}, 46

\bibitem[\protect\citeauthoryear{{Feng} \& {Soria}}{{Feng} \&
  {Soria}}{2011}]{Feng+Soria11}
{Feng}, H.,  \& {Soria}, R. 2011, New Astronomy Reviews, 55, 166

\bibitem[\protect\citeauthoryear{{Gemini Observatory} \& {AURA}}{{Gemini
  Observatory} \& {AURA}}{2016}]{Gemini}
{Gemini Observatory},  \& {AURA}. 2016, {Gemini IRAF: Data reduction software
  for the Gemini telescopes}, Astrophysics Source Code Library

\bibitem[\protect\citeauthoryear{{Gimeno} et~al.}{{Gimeno}
  et~al.}{2016}]{Gimeno+16}
{Gimeno}, G., et~al. 2016, in \procspie, Vol. 9908, Ground-based and Airborne
  Instrumentation for Astronomy VI, 99082S

\bibitem[\protect\citeauthoryear{{Heil}, {Uttley}, \& {Klein-Wolt}}{{Heil}
  et~al.}{2015a}]{Heil+15a}
{Heil}, L.~M., {Uttley}, P.,  \& {Klein-Wolt}, M. 2015a, \mnras, 448, 3348

\bibitem[\protect\citeauthoryear{{Heil}, {Uttley}, \& {Klein-Wolt}}{{Heil}
  et~al.}{2015b}]{Heil+15b}
{Heil}, L.~M., {Uttley}, P.,  \& {Klein-Wolt}, M. 2015b, \mnras, 448, 3339

\bibitem[\protect\citeauthoryear{{Heil}, {Vaughan}, \& {Roberts}}{{Heil}
  et~al.}{2009}]{Heil+09}
{Heil}, L.~M., {Vaughan}, S.,  \& {Roberts}, T.~P. 2009, \mnras, 397, 1061

\bibitem[\protect\citeauthoryear{{Hill} et~al.}{{Hill} et~al.}{2004}]{Hill+04}
{Hill}, J., et~al. 2004, in APS Meeting Abstracts

\bibitem[\protect\citeauthoryear{{Hook} et~al.}{{Hook} et~al.}{2004}]{Hook+04}
{Hook}, I.~M., {J{\o}rgensen}, I., {Allington-Smith}, J.~R., {Davies}, R.~L.,
  {Metcalfe}, N., {Murowinski}, R.~G.,  \& {Crampton}, D. 2004, \pasp, 116, 425

\bibitem[\protect\citeauthoryear{{Immler}, {Brown}, \& {Russell}}{{Immler}
  et~al.}{2010}]{Immler+10}
{Immler}, S., {Brown}, P.,  \& {Russell}, B.~R. 2010, The Astronomer's
  Telegram, 2639, 1

\bibitem[\protect\citeauthoryear{{Ingram} \& {Done}}{{Ingram} \&
  {Done}}{2012}]{Ingram+12}
{Ingram}, A.,  \& {Done}, C. 2012, \mnras, 419, 2369

\bibitem[\protect\citeauthoryear{{Kalberla} et~al.}{{Kalberla}
  et~al.}{2005}]{Kalberla+05}
{Kalberla}, P.~M.~W., {Burton}, W.~B., {Hartmann}, D., {Arnal}, E.~M.,
  {Bajaja}, E., {Morras}, R.,  \& {P{\"o}ppel}, W.~G.~L. 2005, \aap, 440, 775

\bibitem[\protect\citeauthoryear{{Kallman} \& {McCray}}{{Kallman} \&
  {McCray}}{1982}]{Kallman+McCray82}
{Kallman}, T.~R.,  \& {McCray}, R. 1982, \apjs, 50, 263

\bibitem[\protect\citeauthoryear{{Kewley} et~al.}{{Kewley}
  et~al.}{2001}]{Kewley+01}
{Kewley}, L.~J., {Heisler}, C.~A., {Dopita}, M.~A.,  \& {Lumsden}, S. 2001,
  \apjs, 132, 37

\bibitem[\protect\citeauthoryear{{Khan} et~al.}{{Khan} et~al.}{2010}]{Khan+10}
{Khan}, R., {Stanek}, K.~Z., {Kochanek}, C.~S., {Thompson}, T.~A.,  \&
  {Prieto}, J.~L. 2010, The Astronomer's Telegram, 2632, 1

\bibitem[\protect\citeauthoryear{{King} \& {Lasota}}{{King} \&
  {Lasota}}{2016}]{King+16}
{King}, A.,  \& {Lasota}, J.-P. 2016, \mnras, 458, L10

\bibitem[\protect\citeauthoryear{{King}}{{King}}{2004}]{King04}
{King}, A.~R. 2004, Nuclear Physics B Proceedings Supplements, 132, 376

\bibitem[\protect\citeauthoryear{{King}}{{King}}{2009}]{King09}
{King}, A.~R. 2009, \mnras, 393, L41

\bibitem[\protect\citeauthoryear{{Kosec} et~al.}{{Kosec}
  et~al.}{2018}]{Kosec+18}
{Kosec}, P., {Pinto}, C., {Walton}, D.~J., {Fabian}, A.~C., {Bachetti}, M.,
  {F{\"u}rst}, F.,  \& {Grefenstette}, B.~W. 2018, ArXiv e-prints

\bibitem[\protect\citeauthoryear{{Lau} et~al.}{{Lau} et~al.}{2016}]{Lau+16}
{Lau}, R.~M., et~al. 2016, \apj, 830, 142

\bibitem[\protect\citeauthoryear{{Linden} et~al.}{{Linden}
  et~al.}{2010}]{Linden+10}
{Linden}, T., {Kalogera}, V., {Sepinsky}, J.~F., {Prestwich}, A., {Zezas}, A.,
  \& {Gallagher}, J.~S. 2010, \apj, 725, 1984

\bibitem[\protect\citeauthoryear{{Lyubarskii}}{{Lyubarskii}}{1997}]{Lyubarskii97}
{Lyubarskii}, Y.~E. 1997, \mnras, 292, 679

\bibitem[\protect\citeauthoryear{{Middleton} et~al.}{{Middleton}
  et~al.}{2015}]{Middleton+15a}
{Middleton}, M.~J., {Heil}, L., {Pintore}, F., {Walton}, D.~J.,  \& {Roberts},
  T.~P. 2015, \mnras, 447, 3243

\bibitem[\protect\citeauthoryear{{Middleton} et~al.}{{Middleton}
  et~al.}{2011}]{Middleton+11a}
{Middleton}, M.~J., {Roberts}, T.~P., {Done}, C.,  \& {Jackson}, F.~E. 2011,
  \mnras, 411, 644

\bibitem[\protect\citeauthoryear{{Middleton} et~al.}{{Middleton}
  et~al.}{2014}]{Middleton+14a}
{Middleton}, M.~J., {Walton}, D.~J., {Roberts}, T.~P.,  \& {Heil}, L. 2014,
  \mnras, 438, L51

\bibitem[\protect\citeauthoryear{{Monard}}{{Monard}}{2010}]{Monard10}
{Monard}, L.~A.~G. 2010, Central Bureau Electronic Telegrams, 2289, 1

\bibitem[\protect\citeauthoryear{{Pakull} \& {Angebault}}{{Pakull} \&
  {Angebault}}{1986}]{Pakull+Angebault86}
{Pakull}, M.~W.,  \& {Angebault}, L.~P. 1986, \nat, 322, 511

\bibitem[\protect\citeauthoryear{{Pakull}, {Gris{\'e}}, \& {Motch}}{{Pakull}
  et~al.}{2006}]{Pakull+06}
{Pakull}, M.~W., {Gris{\'e}}, F.,  \& {Motch}, C. 2006, in IAU Symposium, Vol.
  230, Populations of High Energy Sources in Galaxies, ed. E.~J.~A. {Meurs} \&
  G.~{Fabbiano}, 293

\bibitem[\protect\citeauthoryear{{Pakull} \& {Mirioni}}{{Pakull} \&
  {Mirioni}}{2002}]{Pakull+Mirioni02}
{Pakull}, M.~W.,  \& {Mirioni}, L. 2002, ArXiv Astrophysics e-prints

\bibitem[\protect\citeauthoryear{{Pakull} \& {Mirioni}}{{Pakull} \&
  {Mirioni}}{2003}]{Pakull+Mirioni03}
{Pakull}, M.~W.,  \& {Mirioni}, L. 2003, in Revista Mexicana de Astronomia y
  Astrofisica Conference Series, Vol.~15, Revista Mexicana de Astronomia y
  Astrofisica Conference Series, ed. J.~{Arthur} \& W.~J. {Henney}, 197

\bibitem[\protect\citeauthoryear{{Poutanen} et~al.}{{Poutanen}
  et~al.}{2013}]{Poutanen+13}
{Poutanen}, J., {Fabrika}, S., {Valeev}, A.~F., {Sholukhova}, O.,  \&
  {Greiner}, J. 2013, \mnras, 432, 506

\bibitem[\protect\citeauthoryear{{Poutanen} et~al.}{{Poutanen}
  et~al.}{2007}]{Poutanen+07}
{Poutanen}, J., {Lipunova}, G., {Fabrika}, S., {Butkevich}, A.~G.,  \&
  {Abolmasov}, P. 2007, \mnras, 377, 1187

\bibitem[\protect\citeauthoryear{{Prieto} et~al.}{{Prieto}
  et~al.}{2010}]{Prieto+10}
{Prieto}, J.~L., {Bond}, H.~E., {Kochanek}, C.~S., {Khan}, R., {Stanek}, K.~Z.,
   \& {Thompson}, T.~A. 2010, The Astronomer's Telegram, 2660, 1

\bibitem[\protect\citeauthoryear{{Remillard} \& {McClintock}}{{Remillard} \&
  {McClintock}}{2006}]{Remillard+06}
{Remillard}, R.~A.,  \& {McClintock}, J.~E. 2006, \araa, 44, 49

\bibitem[\protect\citeauthoryear{{Schlafly} \& {Finkbeiner}}{{Schlafly} \&
  {Finkbeiner}}{2011}]{Schlafly+11}
{Schlafly}, E.~F.,  \& {Finkbeiner}, D.~P. 2011, \apj, 737, 103

\bibitem[\protect\citeauthoryear{{Shakura} \& {Sunyaev}}{{Shakura} \&
  {Sunyaev}}{1973}]{Shakura+73}
{Shakura}, N.~I.,  \& {Sunyaev}, R.~A. 1973, \aap, 24, 337

\bibitem[\protect\citeauthoryear{{Sutherland} \& {Dopita}}{{Sutherland} \&
  {Dopita}}{1993}]{Sutherland+93}
{Sutherland}, R.~S.,  \& {Dopita}, M.~A. 1993, \apjs, 88, 253

\bibitem[\protect\citeauthoryear{{Sutton}, {Roberts}, \& {Middleton}}{{Sutton}
  et~al.}{2013}]{Sutton+13}
{Sutton}, A.~D., {Roberts}, T.~P.,  \& {Middleton}, M.~J. 2013, \mnras, 435,
  1758

\bibitem[\protect\citeauthoryear{{Takeuchi}, {Ohsuga}, \&
  {Mineshige}}{{Takeuchi} et~al.}{2013}]{Takeuchi+13}
{Takeuchi}, S., {Ohsuga}, K.,  \& {Mineshige}, S. 2013, \pasj, 65, 88

\bibitem[\protect\citeauthoryear{{Takeuchi}, {Ohsuga}, \&
  {Mineshige}}{{Takeuchi} et~al.}{2014}]{Takeuchi+14}
{Takeuchi}, S., {Ohsuga}, K.,  \& {Mineshige}, S. 2014, \pasj, 66, 48

\bibitem[\protect\citeauthoryear{{Tarter}, {Tucker}, \& {Salpeter}}{{Tarter}
  et~al.}{1969}]{Tarter+69}
{Tarter}, C.~B., {Tucker}, W.~H.,  \& {Salpeter}, E.~E. 1969, \apj, 156, 943

\bibitem[\protect\citeauthoryear{{van der Klis}}{{van der
  Klis}}{1989}]{vanderKlis89}
{van der Klis}, M. 1989, \araa, 27, 517

\bibitem[\protect\citeauthoryear{{Villar} et~al.}{{Villar}
  et~al.}{2016}]{Villar+16}
{Villar}, V.~A., et~al. 2016, \apj, 830, 11

\bibitem[\protect\citeauthoryear{{Voss} et~al.}{{Voss} et~al.}{2011}]{Voss+11}
{Voss}, R., {Nielsen}, M.~T.~B., {Nelemans}, G., {Fraser}, M.,  \& {Smartt},
  S.~J. 2011, \mnras, 418, L124

\bibitem[\protect\citeauthoryear{{Walton} et~al.}{{Walton}
  et~al.}{2018}]{Walton+18}
{Walton}, D.~J., et~al. 2018, \apjl, 857, L3

\bibitem[\protect\citeauthoryear{{Wiktorowicz} et~al.}{{Wiktorowicz}
  et~al.}{2017}]{Wiktorowicz+17}
{Wiktorowicz}, G., {Sobolewska}, M., {Lasota}, J.-P.,  \& {Belczynski}, K.
  2017, \apj, 846, 17

\bibitem[\protect\citeauthoryear{{Yang}, {Feng}, \& {Kaaret}}{{Yang}
  et~al.}{2011}]{Yang+11}
{Yang}, L., {Feng}, H.,  \& {Kaaret}, P. 2011, \apj, 733, 118

\end{thebibliography}

\end{document}